\documentclass[reprint,sort&compress, nobibnotes, floatfix, aps,  prb]{revtex4-2}
\usepackage{hyperref}
\usepackage[vietnamese, english]{babel}
\usepackage{amssymb}
\usepackage{amsmath}
\usepackage{graphicx}
\usepackage{color}
\hypersetup{
     colorlinks   = true,
     citecolor    = blue
}
\usepackage{physics}
\usepackage{float}
\usepackage{bbold}
\usepackage[normalem]{ulem}
\renewcommand{\vec}[1]{\boldsymbol{#1}}
\graphicspath{{Figs/}}
\newcommand{\be}{\begin{equation}}
\newcommand{\ee}{\end{equation}}
\newcommand{\bea}{\begin{eqnarray}}
\newcommand{\eea}{\end{eqnarray}}

\def\nn{\nonumber}

\def\pref{\eqref}

\begin{document}

\title{Superconductivity induced by the inter-valley Coulomb scattering in a few layers of graphene}

\author{Tommaso Cea}
\affiliation{
Department of Physical and Chemical Sciences,
Universit\'a degli Studi dell'Aquila, I-67100 L'Aquila, Italy
}


\begin{abstract}
We study the inter-valley scattering induced by the Coulomb repulsion as
a purely electronic mechanism for the origin of superconductivity in few layers of graphene.
The pairing is strongly favored by the presence
of van Hove singularities (VHS's) in the density of states (DOS).
We consider three different hetherostructures: 
twisted bilayer graphene (TBG),
rhombohedral trilayer graphene (RTG) and
Bernal bilayer graphene (BBG).
We obtain trends and estimates of the superconducting
(SC) critical temperature in agreement with the experimental findings,
which might identify the inter-valley Coulomb scattering as a
universal pairing mechanism in few layers of graphene.

\end{abstract}

\maketitle

\section{Introduction}
The discovery of superconductivity in TBG\cite{Cao_nature18,Lu_nature19,Yankowitz_science19}
led the scientific community
to a renewed interest in the study of the SC properties of graphene,
that has been further motivated by the more recent observations of
SC behavior in other heterostructures based on graphene:
twisted trilayer graphene (TTG)\cite{Park_nature21},
untwisted RTG\cite{Zhou_nature21} and
BBG\cite{Zhou_science22} in a perpendicular electric field.
In all these systems the SC transition can be
controlled by experimentally tunable parameters, like eg:
the relative twist between the layers, the electronic density and the applied displacement field.
Even though the critical temperatures, $T_c$,
observed so far in these materials do not exceed the scale of a few Kelvin,
the large ratios between $T_c$ and the Fermi energy, up to $\sim10\%$,
suggests that
a strong pairing interaction is at play.
On the other hand, the complex phase diagrams reported in the literature
clearly highlight the strongly correlated behavior.
The recent observation of the Ref. \cite{Zhang_cm22},
that the value of $T_c$ in BBG can be increased by one order of magnitude
by a substrate of WSe$_2$,
emphasizes the highly tunable nature of the pairing,
paving the way towards engineering new techniques
for controlling the magnitude of $T_c$.
Furthermore, the violation of the Pauli's limit reported
in the experiments \cite{Park_nature21,Zhou_nature21,Zhou_science22}
suggests that spin-triplet Cooper pairs are favored in these systems.

On the theoretical side, it is universally accepted that the band flattening and
the vicinity of the VHS's to the Fermi level enhance
the role of the electronic interactions in TBG, TTG, RTG and BBG,
favoring the formation of symmetry broken phases
(see eg the Refs.
\cite{Guinea_pnas18,Sherkunov_prb18,Cea_prb19,
Cea_prb20,Lin_prb20,Chichinadze_cm22}).
However, the debate on the mechanism
at the origin of the superconductivity in these systems is still open.
Many models have been studied so far,
that either consider the superconductivity driven
by purely electronic interactions\cite{Isobe_prx18,Sherkunov_prb18,
Liu_prl18,Gonzalez_prl19,You_npj19,Roy_prb19,Sharma_prr20,
Chichinadze_prb20,Lin_prb20,Qin_cm21,
Dong_cm21,Ghazaryan_prl21,Dai_prl21,Gonzalez_cm21,
Fischer_npj22,Szabo_prb22,Cea_prb22,Qin_cm22,
Crepel_prb22,Lu_prb22,Jimeno_cm22} or
by more conventional phononic mechanisms\cite{Wu_prl18,Peltonen_prb18,Choi_prb18,
Lian_prl19,Angeli_prx19,Wu_prb19,
Schrodi_prr20,Choi_prl21,Chou_prl21,
Chou_prb22,Firoz_cm22}.
The combined effects of the screened Coulomb interaction,
the electronic Umklapp processes and the electron-phonon coupling
have been shown to favor the pairing in
TBG\cite{Lewandowski_prb21,Lewandowski_npj21,Cea_pnas21}
and in TTG\cite{phong_prb21}.
Furthermore, 
the Refs. \cite{Po_prx18,Kozii_cm20,Chatterjee_natcomm22,
Dong_cm21b,Dong_cm22}
explored other unconventional mechanisms, in which the pairing is mediated
by soft electronic collective modes.
Remarkably,
there is not yet a general agreement on whether
the superconductivity observed in TBG and in TTG
has the same origin as in the untwisted RTG and BBG.

In this article, we study the inter-valley scattering
induced by the Coulomb interaction
as a purely electronic mechanism for the origin
of superconductivity in few layers of graphene.
The resulting Cooper pairs are spin-triplets
with the two electrons in opposite valleys, $K,K'$,
featuring $p$- or $f$-wave symmetry.
Because the large momentum transfer, $\Delta K\equiv K-K'$,
involved in the process makes the interaction strength negligible,
a high DOS is necessary to boost the pairing.
This condition is often realized in few layers of graphene,
where the electronic bands can be flattened
by tuning a number of experimental parameters,
thus giving rise to VHS's.
At first approximation, we neglect the contribution of the intra-valley Coulomb repulsion,
which is long-ranged,
since it is drastically screened in the van Hove scenario.
We show quantitatively that this assumption is fully justified in the SI\cite{SI}.
Using effective continuum models with realistic parameters,
we characterize the SC transition induced by the
inter-valley Coulomb scattering in
TBG, RTG and BBG, upon varying the relative twist between the layers and/or
the electronic density and/or the displacement field.
We obtain estimates and trends of $T_c$
in good agreement with the experimental results,
emphasizing the strong enhancement of $T_c$
by the presence of VHS's.
Remarkably, our calculations account for the different orders of magnitude
of the critical temperatures observed in different materials.
Considering also the experimental evidence
of spin-triplet superconductivity in these systems,
our study might identify the inter-valley Coulomb scattering as a universal driving mechanism
for the superconductivity observed so far in few layers of graphene.
We also identify a non-trivial structure of the SC order parameter (OP) in real space.


\section{The model: effective attraction from the inter-valley scattering}
Our theoretical description of the pairing interaction starts from considering the
Coulomb repulsion between the $p_z$
electrons within the minimal lattice model for a multilayer of graphene:
\begin{widetext}
\bea\label{H_coulomb}
\hat{H}_{int}=\frac{1}{2}
\sum_{\vec{R}\vec{R}'}
\sum_{ij\sigma\sigma'}
c^{\dagger}_{i\sigma}(\vec{R})c^{\dagger}_{j\sigma'}(\vec{R}')
V^{ij}_{C}(\vec{R}-\vec{R}')
c_{j\sigma'}(\vec{R}')c_{i\sigma}(\vec{R}),
\eea
\end{widetext}
where $\vec{R}$ are the coordinates of the Bravais lattice, $i,j$ are the labels of the sub-lattice/layer,
$c_{i\sigma}(\vec{R})$ is the the quantum operator for the annihilation
of one electron with spin $\sigma$ in the $p_z$ orbital localized at the position $\vec{R}+\vec{\delta}_i$,
$\vec{\delta}_i$ being the internal coordinate in the unit cell,
and:
\bea
V^{ij}_C(\vec{R}-\vec{R}')=\frac{e^2}{4\pi\epsilon\left| 
\vec{R}-\vec{R}'+\vec{\delta}_i-\vec{\delta}_j
\right|}
\eea
is the Coulomb potential, where $e$ is the electron charge and
$\epsilon$ is the dielectric constant of the environment,
$\epsilon=\epsilon_0$ in the vacuum.
Next, we consider the continuum limit of the lattice model,
by expanding the operators $c$ as:
\bea\label{continuum_fields}
c_{i\sigma}(\vec{R})\equiv A_c^{1/2}\left[
\psi^{K}_{i\sigma}(\vec{R})e^{iK\cdot\vec{R}}+
\psi^{K'}_{i\sigma}(\vec{R})e^{iK'\cdot\vec{R}}
\right],
\eea
where $A_c=\sqrt{3}a^2/2$ is the area of the unit cell of graphene,
$a=2.46${\AA} being the lattice constant,
 $K,K'$ are the non equivalent corners of the BZ
and $\psi^{K}_{i\sigma}(\vec{r}),\psi^{K'}_{i\sigma}(\vec{r})$
are fermionic operators, which vary smoothly with the continuum position, $\vec{r}$,
and represent the valley projections of $c_{i\sigma}(\vec{R})$.
Replacing the Eq. \pref{continuum_fields} into the Eq. \pref{H_coulomb},
among all the terms one finds the following valley-exchange interaction:
\begin{widetext}
\bea\label{Hexc}
\hat{H}_{exc}=
A_c^2
\sum_{\vec{R}\vec{R}'}
\sum_{ij\sigma\sigma'}
\psi^{K,\dagger}_{i\sigma}(\vec{R})\psi^{K',\dagger}_{j\sigma'}(\vec{R}')
\psi^{K}_{j\sigma'}(\vec{R}')\psi^{K'}_{i\sigma}(\vec{R})
V^{ij}_{C}(\vec{R}-\vec{R}')e^{-i\Delta K\cdot(\vec{R}-\vec{R}')},
\eea
\end{widetext}
which describes the inter-valley scattering processes.
As we show in detail in the supplementary information (SI)\cite{SI},
the Eq. \pref{Hexc} can be safely approximated by the continuum Hamiltonian:
\bea
\hat{H}_{exc}&\simeq&\nn\\
-J\sum_{i\sigma\sigma'}
\int&\,d^2\vec{r}&
\psi^{K,\dagger}_{i\sigma}(\vec{r})\psi^{K',\dagger}_{i\sigma'}(\vec{r})
\psi^{K'}_{i\sigma}(\vec{r})\psi^{K}_{i\sigma'}(\vec{r}),\label{Hexc_cont}
\eea
where: $J\equiv\frac{e^2}{2\epsilon|\Delta K|}$
is the Fourier transform of the Coulomb potential in 2D, evaluated at $\Delta K$.
It's worth noting that the interaction described by the Eq. \pref{Hexc_cont} is purely local,
not only in the space coordinates, but also in the sub-lattice and layer indices.
Because $J>0$, $\hat{H}_{exc}$
provides an effective attraction in the spin-triplet channel (see the SI\cite{SI}),
favoring the Cooper pairing with electrons in opposite valleys.
This kind of interaction belongs to the universality class identified by
Crepel and Fu\cite{crepel_scadv22,crepel_pnas22},
who have demonstrated the relevance of
the valley-exchange interaction in inducing the pairing in narrow band systems. 
As we already mentioned,
we stress that the spin triplet superconductivity is
a general claim of the experimental works.
On the other hand, the valley-exchange from
the Coulomb interaction has been shown to favor spin-triplet superconductivity
in various materials
(see for example the Refs. \cite{GuineaUchoa_prb12,Roldan_prb13,Qin_cm22}).
The SC OP, $\Delta^i(\vec{r})$, is purely local and
the value of $T_c$ can be obtained within the BCS theory
as the largest temperature for which it exists a
nonzero solution of the linearized gap equation:
\bea\label{linearized_gap_eq}
\Delta^i(\vec{r})&=&
\frac{J}{\beta}\sum_j\int\,d\vec{r}'\sum_{l=-\infty}^{+\infty}\times\\
&\times&\mathcal{G}^K_{ij}\left(\vec{r},\vec{r}';i\omega_l\right)
\mathcal{G}^{K'}_{ij}\left(\vec{r},\vec{r}';-i\omega_l\right)
\Delta^j\left(\vec{r}'\right),\nn
\eea 
where $\beta=(K_BT)^{-1}$ is the inverse of the temperature,
$\omega_l=\pi (2l+1)/(\hbar\beta)$ are fermionic Matsubara frequencies
and $\mathcal{G}^{K,K'}$ are the Green's function for the $K,K'$ valleys,
respectively, computed in the normal phase.
The Eq. \pref{linearized_gap_eq} is written in real space in order to be as general as possible,
holding also for non-translationally invariant systems,
as is the case of the TBG that we will consider below.

It's worth noting that:
i) $J$ is generally small. For example, if we consider $\epsilon/\epsilon_0=4$,
which mimics the screening by a substrate of hBN, then $J\simeq 13.25$eV{\AA}$^2$,
consistent with the estimates of the Hubbard interaction strength in
graphene\cite{Parr_JCP1950,Ohno_TCA1964,Baird_JCP1969,
Verges_prb10,Wehling_prl11}.
Such a small value requires a large DOS at the Fermi energy, $N_F$,
for making the dimensionless SC coupling, $\lambda=N_FJ$, sizeable.
While the DOS is suppressed in the monolayer graphene close to charge neutrality,
multilayer stacks of graphene offer a way to increase the value of $N_F$,
and hence to strengthen $\lambda$, upon tuning a number of experimental parameters,
like eg the relative twist between the layers, the electronic density, the displacement field etc.;
ii) we are not considering the effects of the intra-valley scattering
induced by the Coulomb interaction at small momenta, which are repulsive.
In a van Hove scenario, these terms are suppressed by the strong internal screening.
As we show in the SI\cite{SI} for the case of the RTG,
the strength of the screened intra-valley Coulomb repulsion is orders of magnitude smaller
than $J$, which fully justifies its omission;
iii) we are not considering the internal screening of $J$,
which is supposed to be negligible as it is induced by the particle-hole excitations
with the particle and the hole in opposite valleys.
These kind of processes are indeed suppressed despite of a large DOS.
This assumption is justified quantitatively in the SI\cite{SI},
where we show that the screening essentially does not affect the value of $J$ as compared to its bare value.
\section{Results}
\textbf{\textit{The case of the TBG.}}
A relative small twist, $\theta$, between the two layers of a bilayer graphene
generates a moir\'e superlattice with periodicity:
$L_m\simeq a/\theta$, much larger than the lattice constant of the monolayer graphene.
The inter-layer hopping varies smoothly at the scale of the moir\'e,
breaking the translational invariance within each moir\'e unit cell and
strongly hybridizing the $p_z$
orbitals of the constitutive graphene sheets.
The superconductivity has been observed at the ''magic'' angle,
$\theta=1.05^\circ$\cite{Cao_nature18,Lu_nature19,Yankowitz_science19},
where the free electron spectrum features two weakly dispersing narrow bands
at the charge neutrality point (CNP),
generating strong VHS's
in the DOS\cite{dossantos_prl07,bistritzer_pnas11}.

\begin{figure}
\centering
\includegraphics[width=\columnwidth]{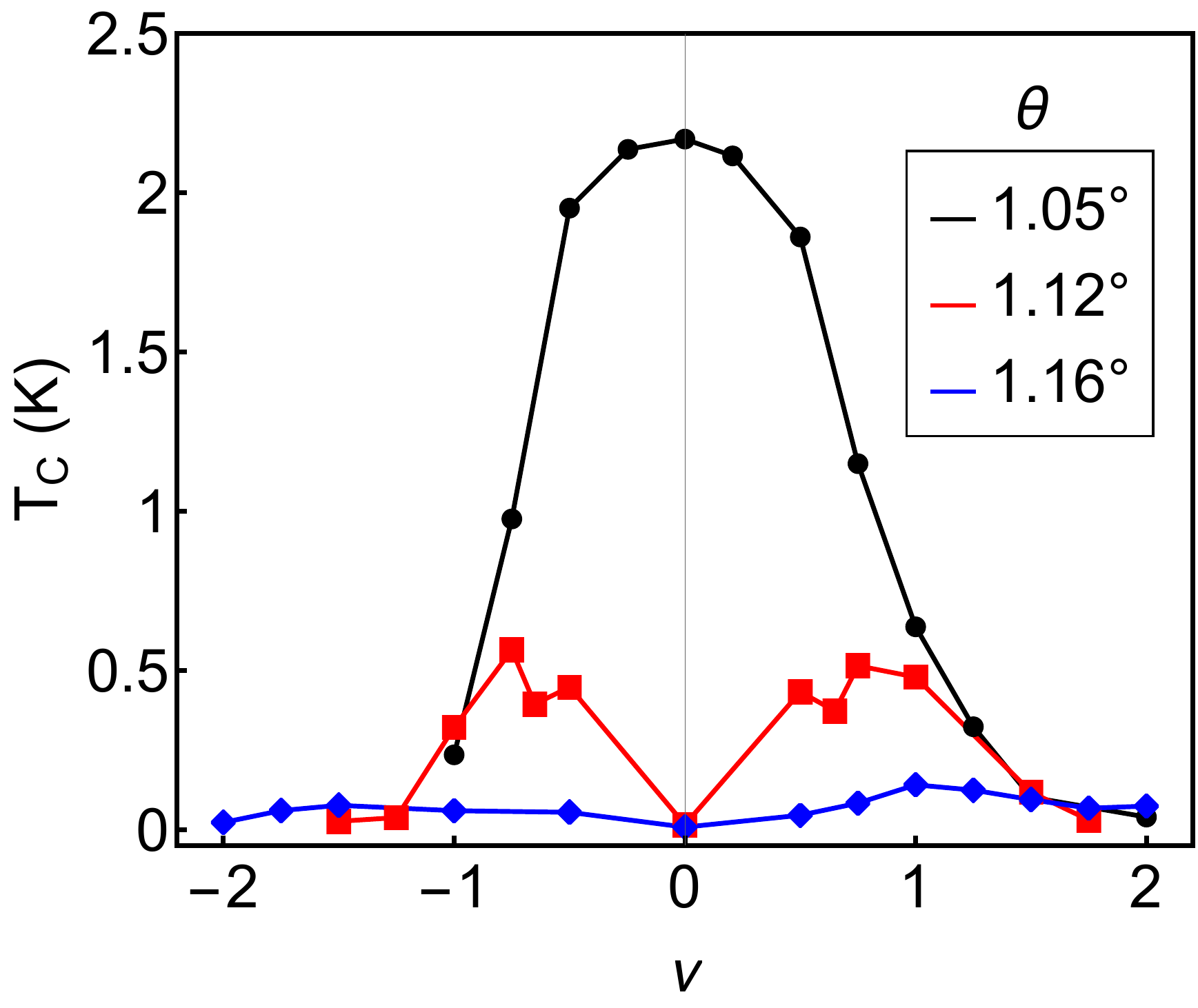}
\caption{
$T_c$ as a function of the filling, obtained for:
$\epsilon/\epsilon_0=4$ and $\theta=1.05^\circ,1.12^\circ,1.16^\circ$.
}
\label{fig:TBG_TC_angle_dep}
\end{figure}

\begin{figure}
\centering
\includegraphics[width=\columnwidth]{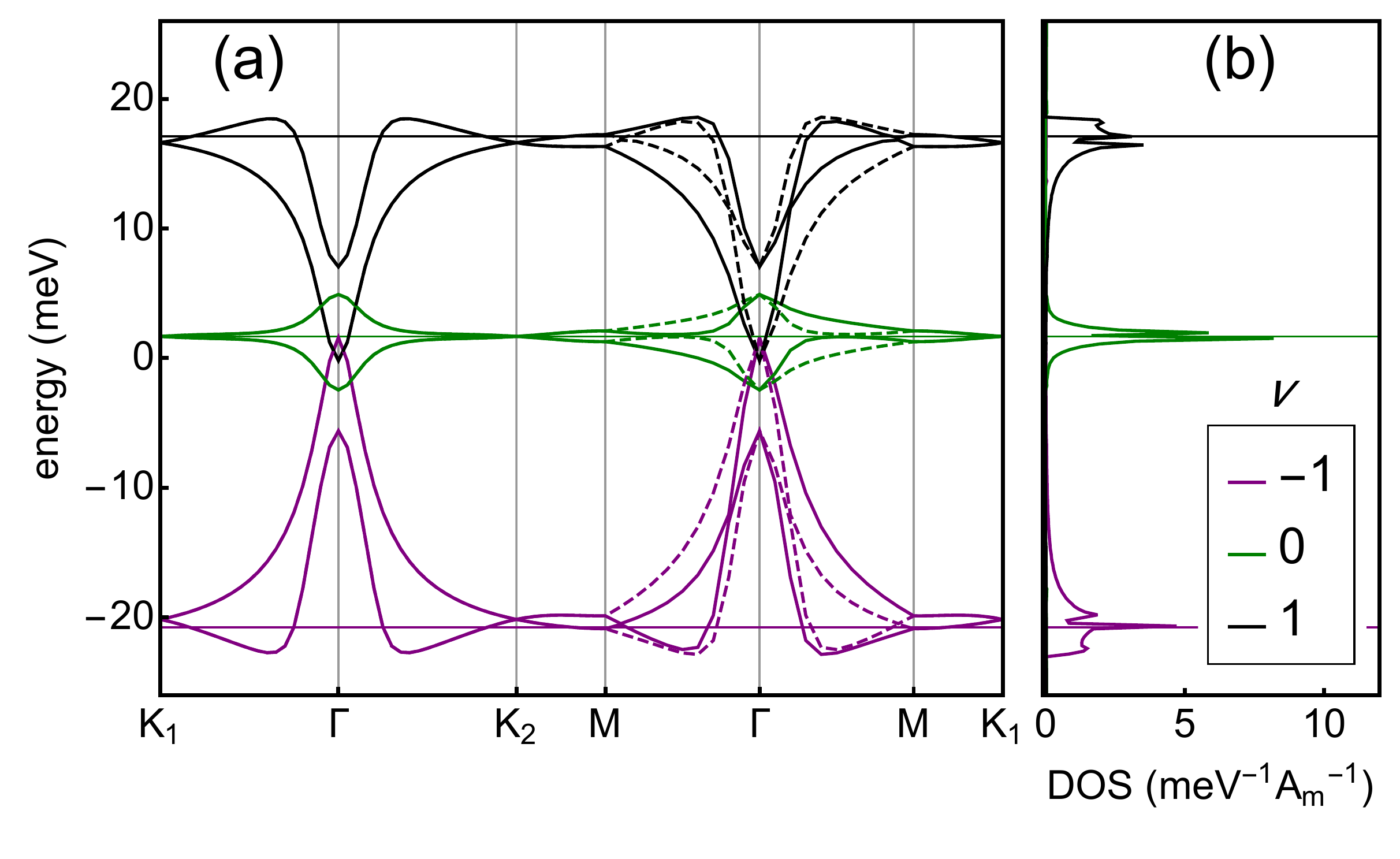}
\caption{
Band structure (a) and DOS (b) corresponding to the two central bands of the TBG,
obtained for:
$\theta=1.05^\circ$ and $\nu=-1,0,1$.
The continuous and the dashed lines refer to the $K$ and $K'$ valleys, respectively,
while the horizontal lines identify the Fermi energies.
The DOS is expressed in units of meV$^{-1}A_m^{-1}$,
where $A_m=\sqrt{3}L^2_m/2$ is the area of the moir\'e unit cell.
}
\label{fig:TBG_1p05deg_bands}
\end{figure}

Using the continuum model of the
TBG\cite{dossantos_prl07,bistritzer_pnas11,dossantos_prb12,koshino_prx18},
we solve the linearized gap equation \pref{linearized_gap_eq} as detailed in the SI\cite{SI}.
The Fig. \ref{fig:TBG_TC_angle_dep} shows
$T_c$ as a function of the filling per moir\'e unit cell, $\nu$,
 obtained for three values of the twist angle,
 as coded in the caption, and for $\epsilon/\epsilon_0=4$.
We find values of $T_c$ of the order of 1K,
in good agreement with the experimental findings.
$T_c$ is the largest for $\theta=1.05^\circ$,
where the bandwidth at the CNP is minimum.
The band structure and the DOS
corresponding to the two central bands
of the TBG at $\theta=1.05^\circ$ are shown
in the Fig. \ref{fig:TBG_1p05deg_bands} for: $\nu=-1,0,1$.
The continuous and the dashed lines refer to the $K$ and $K'$ valleys, respectively,
while the horizontal lines identify the Fermi energies.
Note that the reshaping of the bands with the filling is induced by the
Hartree corrections\cite{Guinea_pnas18,Cea_prb19}.
Comparing the Figs. \ref{fig:TBG_TC_angle_dep}
and \ref{fig:TBG_1p05deg_bands} makes it clear that
the value of $T_c$ increases with $N_F$,
which explains why we obtain the largest $T_c$ at $\nu=0$,
where the bandwidth is minimum.
It's worth noting that the TBG is actually a flavor polarized insulator
at $\nu=0$.
As we are not
considering flavor polarized phases, the superconductivity that we obtain at $\nu=0$ is not realistic.
Nonetheless, it serves as an example to grab the relevance of the VHS's in inducing the pairing.
As we show in the SI\cite{SI},
the order of magnitude of $T_c$ and its $\nu$-dependence
do not change for realistic values of the external screening:
$\epsilon/\epsilon_0\sim 4-6$.

The OP, $\Delta^i(\vec{r})$, that we obtain as solution of
the Eq. \pref{linearized_gap_eq} at $T=T_c$, is almost uniform in the sub-lattices and layers,
has a constant phase and varies in the moir\'e unit cell,
reaching its maxima in the regions with local $AA$ stacking.
Maps of $\Delta^i(\vec{r})$ are shown in the SI\cite{SI}.

\textbf{\textit{The case of the RTG and BBG.}}
Both RTG and BBG become superconductors at very low electronic densities:
$n_e\sim10^{12}$cm$^{-2}$, when an electric field is applied
perpendicular to the graphene's flakes\cite{Zhou_nature21,Zhou_science22}.
The experimental $T_c$'s are $\sim 10^{-1}$K for the RTG and $\sim 10^{-2}$K for the BBG.
From the electronic point of view,
a perpendicular electric field breaks the inversion symmetry and gaps out the Dirac points
in both RTG and BBG.
The nearly flat dispersion close to the gap's edge gives rise to a pronounced VHS.
For values of $n_e$ close to those at which the superconductivity has been reported,
the Fermi level lies near the VHS.

\begin{figure}
\centering
\includegraphics[width=\columnwidth]{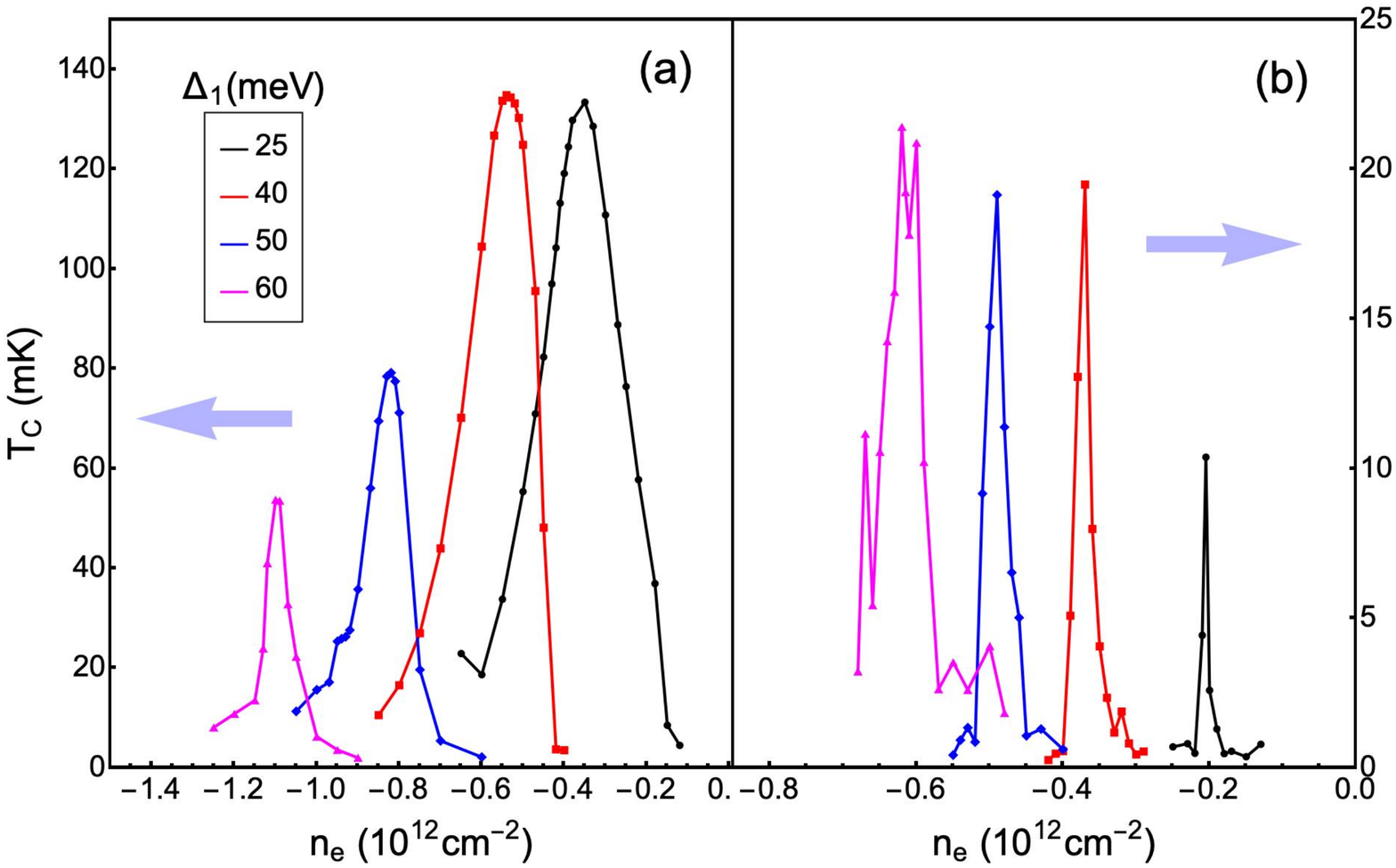}
\caption{
$T_c$ of the hole-doped RTG (a) and BBG (b)
as a function of $n_e$,
obtained for: $\epsilon/\epsilon_0=4$ and $\Delta_1$
as coded in the inset panel.
}
\label{fig:RTG_and_BBG_TC}
\end{figure}

\begin{figure}
\centering
\includegraphics[width=\columnwidth]{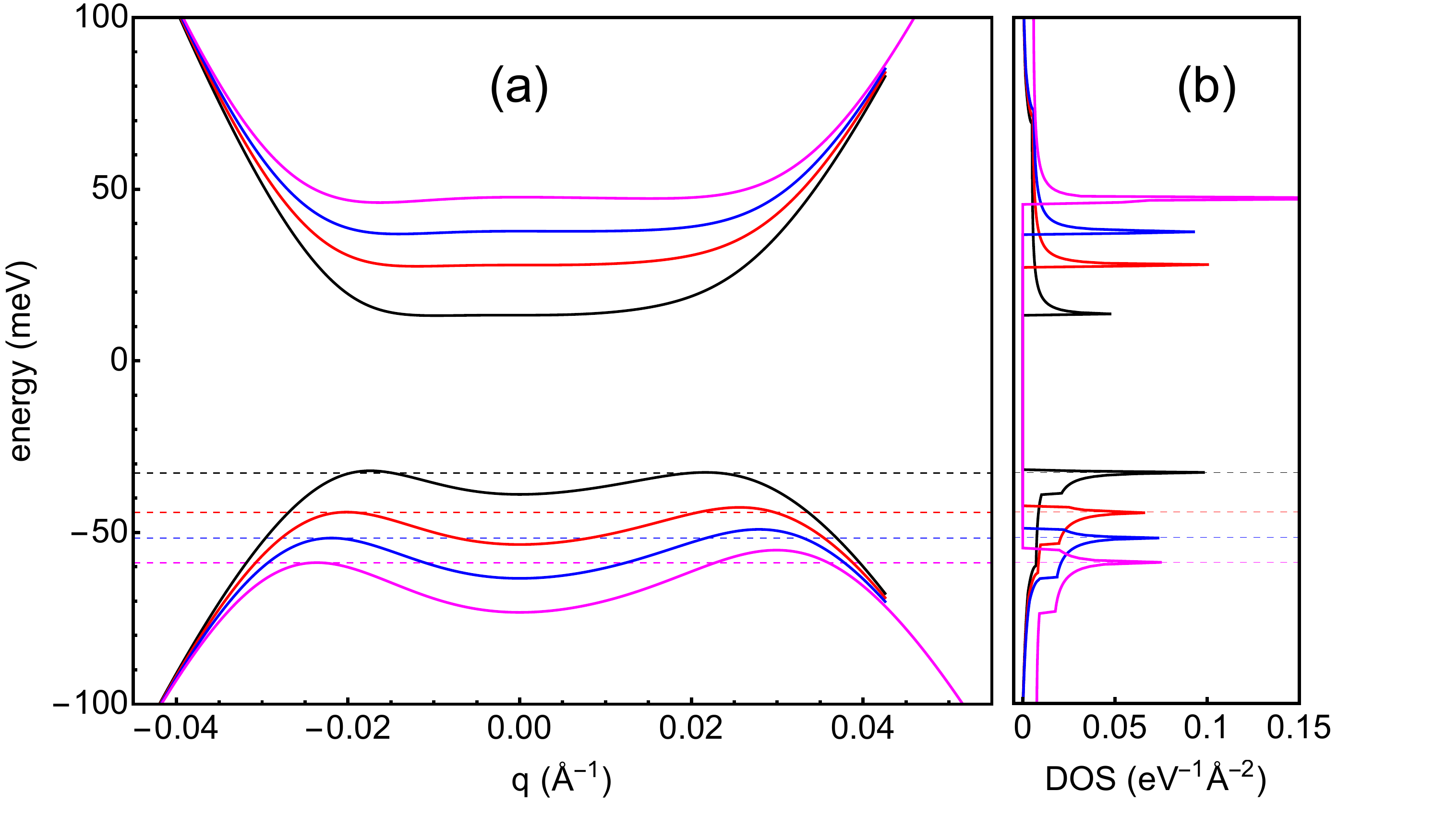}
\caption{
Band structure (a) and DOS (b) of the RTG close to the CNP,
obtained for: $\Delta_1=25,40,50,60$meV, and color coded as in the
Fig. \ref{fig:RTG_and_BBG_TC}.
Here $q=0$ at the $K$ point.
The horizontal dashed lines indicate the Fermi levels
corresponding to the densities that maximize $T_c$ in the
Fig. \ref{fig:RTG_and_BBG_TC}.
}
\label{fig:RTG_bands}
\end{figure}

We study the low energy continuum model of the RTG and BBG,
with the hopping amplitudes
given by the Refs. \cite{zibrov_prl18,Zhou_nature21_bis,Cea_prb22},
and an inter-layer bias, $\Delta_1$,
describing the perpendicular electric field (see the SI\cite{SI}).
We extract $T_c$ by using
the linearized gap equation \pref{linearized_gap_eq}
in the translationally invariant case,
as detailed in the SI\cite{SI}.
The Fig. \ref{fig:RTG_and_BBG_TC} shows the values of $T_c$
as a function of $n_e$ in the hole-doped RTG (a) and BBG (b),
obtained for: $\epsilon/\epsilon_0=4$ and
realistic values of $\Delta_1$, as coded in the inset panel.
We obtain critical temperatures up to $\sim130-140$mK for the RTG,
and up to $\sim20$mK for the BBG,
approximately one order of magnitude smaller than in RTG.
The peak's intensity and position depend on $\Delta_1$.
These estimates of $T_c$
are in excellent agreement with the two experiments
\cite{Zhou_nature21,Zhou_science22}.
The Ref. \cite{Zhou_nature21} identifies two distinct SC phases of the RTG,
named SC1 and SC2.
While the former does respect the Pauli's limit, the latter does not,
implying that the pairing is spin-unpolarized in SC1 and spin-polarized in SC2.
The superconductivity observed in the BBG in the Ref. \cite{Zhou_science22} is always accompanied by the violation of the Pauli's limit,
suggesting that the SC1 phase is absent in the BBG.
It's worth noting that the spin-triplet pairing that we are considering is compatible
with both spin-unpolarized and spin-polarized Cooper pairs,
and consequently it does not exclude a priori either the SC1 or the SC2 regime.
However, comparing the values of $T_c$ in the Fig. \ref{fig:RTG_and_BBG_TC}(a)
with the experimental temperatures reported in the Ref. \cite{Zhou_nature21},
it seems that our calculations are missing the SC2 phase of the RTG.
Remarkably, our results reproduce quite well both
the range of $n_e$ in which the superconductivity has been reported
and the narrowness of the SC domain.
The Fig. \ref{fig:RTG_bands} shows the band structure (a) and DOS (b)
of the RTG close to the CNP,
obtained for the values of $\Delta_1$ considered in the Fig. \ref{fig:RTG_and_BBG_TC}
and color coded as there.
The bands flatten close to the gap edge,
giving rise to the VHS's in the DOS,
whose position and intensity can be tuned by $\Delta_1$.
The horizontal dashed lines indicate the Fermi levels
corresponding to the densities that maximize $T_c$ in the
Fig. \ref{fig:RTG_and_BBG_TC},
emphasizing that the superconductivity is strongly favored when
the Fermi level matches the VHS.
The spectral feature of the BBG in the presence
of a perpendicular electric field is qualitatively very similar to that of the RTG.
The narrowness of the SC domain can be explained by considering that the bands of the
RTG and BBG remain rigid upon tuning $n_e$, that only moves the Fermi level.
Thus, there is only a small range of values of $n_e$ for which the
Fermi level lies near the VHS.
This is in stark contrast to the case of the TBG,
where the shape of the bands changes with the filling
in order to pin the VHS to the Fermi level\cite{Guinea_pnas18,Cea_prb19},
then allowing for wide SC domains.

Finally, we find that the SC OP is fully localized in the layer and sub-lattice in which
the electric field confines the low energy electrons,
as shown in the Ref. \cite{Zhang_prb10}.
 
\section{Conclusions}
Our work represents one step further in the study of
the origin of the superconductivity recently observed in few layers of graphene,
that can be explained by a universal purely electronic mechanism,
in which the valley-exchange induced by the Coulomb repulsion
serves as the paring glue in the valley-singlet/spin-triplet Cooper channel.
The pairing is then favored by the proximity of the Fermi level to the VHS's,
that can be achieved in these materials by tuning a set of experimental parameters.
Exploiting low energy continuum models,
we provide a general method for computing the SC
critical temperature as well as the symmetry of the OP,
and we test it in three representative systems: TBG, RTG and BBG,
obtaining an excellent agreement with the experimental findings.
Beyond the case of the graphene-based heterostructures,
our analysis might be generalized to study the pairing in
the wider class of materials displaying
valley degrees of freedom and VHS's,
as is the case of the recently discovered
superconductors on Silicon surface\cite{Nakamura_prb18,Wu_prl20,Ming_cm22}.


\section{Acknowledgements}
We thank Francisco Guinea and Nicol\'o Defenu for useful discussions.
We acknowledge fundings from the European Commision,
under the Graphene Flagship, Core 3, grant number 881603.

\bibliography{Literature}

\begin{thebibliography}{72}%
\makeatletter
\providecommand \@ifxundefined [1]{%
 \@ifx{#1\undefined}
}%
\providecommand \@ifnum [1]{%
 \ifnum #1\expandafter \@firstoftwo
 \else \expandafter \@secondoftwo
 \fi
}%
\providecommand \@ifx [1]{%
 \ifx #1\expandafter \@firstoftwo
 \else \expandafter \@secondoftwo
 \fi
}%
\providecommand \natexlab [1]{#1}%
\providecommand \enquote  [1]{``#1''}%
\providecommand \bibnamefont  [1]{#1}%
\providecommand \bibfnamefont [1]{#1}%
\providecommand \citenamefont [1]{#1}%
\providecommand \href@noop [0]{\@secondoftwo}%
\providecommand \href [0]{\begingroup \@sanitize@url \@href}%
\providecommand \@href[1]{\@@startlink{#1}\@@href}%
\providecommand \@@href[1]{\endgroup#1\@@endlink}%
\providecommand \@sanitize@url [0]{\catcode `\\12\catcode `\$12\catcode
  `\&12\catcode `\#12\catcode `\^12\catcode `\_12\catcode `\%12\relax}%
\providecommand \@@startlink[1]{}%
\providecommand \@@endlink[0]{}%
\providecommand \url  [0]{\begingroup\@sanitize@url \@url }%
\providecommand \@url [1]{\endgroup\@href {#1}{\urlprefix }}%
\providecommand \urlprefix  [0]{URL }%
\providecommand \Eprint [0]{\href }%
\providecommand \doibase [0]{https://doi.org/}%
\providecommand \selectlanguage [0]{\@gobble}%
\providecommand \bibinfo  [0]{\@secondoftwo}%
\providecommand \bibfield  [0]{\@secondoftwo}%
\providecommand \translation [1]{[#1]}%
\providecommand \BibitemOpen [0]{}%
\providecommand \bibitemStop [0]{}%
\providecommand \bibitemNoStop [0]{.\EOS\space}%
\providecommand \EOS [0]{\spacefactor3000\relax}%
\providecommand \BibitemShut  [1]{\csname bibitem#1\endcsname}%
\let\auto@bib@innerbib\@empty
\bibitem [{\citenamefont {Cao}\ \emph {et~al.}(2018)\citenamefont {Cao},
  \citenamefont {Fatemi}, \citenamefont {Fang}, \citenamefont {Watanabe},
  \citenamefont {Taniguchi}, \citenamefont {Kaxiras},\ and\ \citenamefont
  {Jarillo-Herrero}}]{Cao_nature18}%
  \BibitemOpen
  \bibfield  {author} {\bibinfo {author} {\bibfnamefont {Y.}~\bibnamefont
  {Cao}}, \bibinfo {author} {\bibfnamefont {V.}~\bibnamefont {Fatemi}},
  \bibinfo {author} {\bibfnamefont {S.}~\bibnamefont {Fang}}, \bibinfo {author}
  {\bibfnamefont {K.}~\bibnamefont {Watanabe}}, \bibinfo {author}
  {\bibfnamefont {T.}~\bibnamefont {Taniguchi}}, \bibinfo {author}
  {\bibfnamefont {E.}~\bibnamefont {Kaxiras}},\ and\ \bibinfo {author}
  {\bibfnamefont {P.}~\bibnamefont {Jarillo-Herrero}},\ }\bibfield  {title}
  {\bibinfo {title} {Unconventional superconductivity in magic-angle graphene
  superlattices},\ }\href {https://doi.org/10.1038/nature26160} {\bibfield
  {journal} {\bibinfo  {journal} {Nature}\ }\textbf {\bibinfo {volume} {556}},\
  \bibinfo {pages} {43} (\bibinfo {year} {2018})}\BibitemShut {NoStop}%
\bibitem [{\citenamefont {Lu}\ \emph {et~al.}(2019)\citenamefont {Lu},
  \citenamefont {Stepanov}, \citenamefont {Yang}, \citenamefont {Xie},
  \citenamefont {Aamir}, \citenamefont {Das}, \citenamefont {Urgell},
  \citenamefont {Watanabe}, \citenamefont {Taniguchi}, \citenamefont {Zhang},
  \citenamefont {Bachtold}, \citenamefont {MacDonald},\ and\ \citenamefont
  {Efetov}}]{Lu_nature19}%
  \BibitemOpen
  \bibfield  {author} {\bibinfo {author} {\bibfnamefont {X.}~\bibnamefont
  {Lu}}, \bibinfo {author} {\bibfnamefont {P.}~\bibnamefont {Stepanov}},
  \bibinfo {author} {\bibfnamefont {W.}~\bibnamefont {Yang}}, \bibinfo {author}
  {\bibfnamefont {M.}~\bibnamefont {Xie}}, \bibinfo {author} {\bibfnamefont
  {M.~A.}\ \bibnamefont {Aamir}}, \bibinfo {author} {\bibfnamefont
  {I.}~\bibnamefont {Das}}, \bibinfo {author} {\bibfnamefont {C.}~\bibnamefont
  {Urgell}}, \bibinfo {author} {\bibfnamefont {K.}~\bibnamefont {Watanabe}},
  \bibinfo {author} {\bibfnamefont {T.}~\bibnamefont {Taniguchi}}, \bibinfo
  {author} {\bibfnamefont {G.}~\bibnamefont {Zhang}}, \bibinfo {author}
  {\bibfnamefont {A.}~\bibnamefont {Bachtold}}, \bibinfo {author}
  {\bibfnamefont {A.~H.}\ \bibnamefont {MacDonald}},\ and\ \bibinfo {author}
  {\bibfnamefont {D.~K.}\ \bibnamefont {Efetov}},\ }\bibfield  {title}
  {\bibinfo {title} {Superconductors, orbital magnets and correlated states in
  magic-angle bilayer graphene},\ }\href
  {https://doi.org/10.1038/s41586-019-1695-0} {\bibfield  {journal} {\bibinfo
  {journal} {Nature}\ }\textbf {\bibinfo {volume} {574}},\ \bibinfo {pages}
  {653} (\bibinfo {year} {2019})}\BibitemShut {NoStop}%
\bibitem [{\citenamefont {Yankowitz}\ \emph {et~al.}(2019)\citenamefont
  {Yankowitz}, \citenamefont {Chen}, \citenamefont {Polshyn}, \citenamefont
  {Zhang}, \citenamefont {Watanabe}, \citenamefont {Taniguchi}, \citenamefont
  {Graf}, \citenamefont {Young},\ and\ \citenamefont
  {Dean}}]{Yankowitz_science19}%
  \BibitemOpen
  \bibfield  {author} {\bibinfo {author} {\bibfnamefont {M.}~\bibnamefont
  {Yankowitz}}, \bibinfo {author} {\bibfnamefont {S.}~\bibnamefont {Chen}},
  \bibinfo {author} {\bibfnamefont {H.}~\bibnamefont {Polshyn}}, \bibinfo
  {author} {\bibfnamefont {Y.}~\bibnamefont {Zhang}}, \bibinfo {author}
  {\bibfnamefont {K.}~\bibnamefont {Watanabe}}, \bibinfo {author}
  {\bibfnamefont {T.}~\bibnamefont {Taniguchi}}, \bibinfo {author}
  {\bibfnamefont {D.}~\bibnamefont {Graf}}, \bibinfo {author} {\bibfnamefont
  {A.~F.}\ \bibnamefont {Young}},\ and\ \bibinfo {author} {\bibfnamefont
  {C.~R.}\ \bibnamefont {Dean}},\ }\bibfield  {title} {\bibinfo {title} {Tuning
  superconductivity in twisted bilayer graphene},\ }\href
  {https://doi.org/10.1126/science.aav1910} {\bibfield  {journal} {\bibinfo
  {journal} {Science}\ }\textbf {\bibinfo {volume} {363}},\ \bibinfo {pages}
  {1059} (\bibinfo {year} {2019})},\ \Eprint
  {https://arxiv.org/abs/https://science.sciencemag.org/content/363/6431/1059.full.pdf}
  {https://science.sciencemag.org/content/363/6431/1059.full.pdf} \BibitemShut
  {NoStop}%
\bibitem [{\citenamefont {Park}\ \emph {et~al.}(2021)\citenamefont {Park},
  \citenamefont {Cao}, \citenamefont {Watanabe}, \citenamefont {Taniguchi},\
  and\ \citenamefont {Jarillo-Herrero}}]{Park_nature21}%
  \BibitemOpen
  \bibfield  {author} {\bibinfo {author} {\bibfnamefont {J.~M.}\ \bibnamefont
  {Park}}, \bibinfo {author} {\bibfnamefont {Y.}~\bibnamefont {Cao}}, \bibinfo
  {author} {\bibfnamefont {K.}~\bibnamefont {Watanabe}}, \bibinfo {author}
  {\bibfnamefont {T.}~\bibnamefont {Taniguchi}},\ and\ \bibinfo {author}
  {\bibfnamefont {P.}~\bibnamefont {Jarillo-Herrero}},\ }\bibfield  {title}
  {\bibinfo {title} {Tunable strongly coupled superconductivity in magic-angle
  twisted trilayer graphene},\ }\href
  {https://doi.org/10.1038/s41586-021-03192-0} {\bibfield  {journal} {\bibinfo
  {journal} {Nature}\ }\textbf {\bibinfo {volume} {590}},\ \bibinfo {pages}
  {249} (\bibinfo {year} {2021})}\BibitemShut {NoStop}%
\bibitem [{\citenamefont {Zhou}\ \emph
  {et~al.}(2021{\natexlab{a}})\citenamefont {Zhou}, \citenamefont {Xie},
  \citenamefont {Taniguchi}, \citenamefont {Watanabe},\ and\ \citenamefont
  {Young}}]{Zhou_nature21}%
  \BibitemOpen
  \bibfield  {author} {\bibinfo {author} {\bibfnamefont {H.}~\bibnamefont
  {Zhou}}, \bibinfo {author} {\bibfnamefont {T.}~\bibnamefont {Xie}}, \bibinfo
  {author} {\bibfnamefont {T.}~\bibnamefont {Taniguchi}}, \bibinfo {author}
  {\bibfnamefont {K.}~\bibnamefont {Watanabe}},\ and\ \bibinfo {author}
  {\bibfnamefont {A.~F.}\ \bibnamefont {Young}},\ }\bibfield  {title} {\bibinfo
  {title} {Superconductivity in rhombohedral trilayer graphene},\ }\href
  {https://doi.org/10.1038/s41586-021-03926-0} {\bibfield  {journal} {\bibinfo
  {journal} {Nature}\ }\textbf {\bibinfo {volume} {598}},\ \bibinfo {pages}
  {434} (\bibinfo {year} {2021}{\natexlab{a}})}\BibitemShut {NoStop}%
\bibitem [{\citenamefont {Zhou}\ \emph {et~al.}(2022)\citenamefont {Zhou},
  \citenamefont {Holleis}, \citenamefont {Saito}, \citenamefont {Cohen},
  \citenamefont {Huynh}, \citenamefont {Patterson}, \citenamefont {Yang},
  \citenamefont {Taniguchi}, \citenamefont {Watanabe},\ and\ \citenamefont
  {Young}}]{Zhou_science22}%
  \BibitemOpen
  \bibfield  {author} {\bibinfo {author} {\bibfnamefont {H.}~\bibnamefont
  {Zhou}}, \bibinfo {author} {\bibfnamefont {L.}~\bibnamefont {Holleis}},
  \bibinfo {author} {\bibfnamefont {Y.}~\bibnamefont {Saito}}, \bibinfo
  {author} {\bibfnamefont {L.}~\bibnamefont {Cohen}}, \bibinfo {author}
  {\bibfnamefont {W.}~\bibnamefont {Huynh}}, \bibinfo {author} {\bibfnamefont
  {C.~L.}\ \bibnamefont {Patterson}}, \bibinfo {author} {\bibfnamefont
  {F.}~\bibnamefont {Yang}}, \bibinfo {author} {\bibfnamefont {T.}~\bibnamefont
  {Taniguchi}}, \bibinfo {author} {\bibfnamefont {K.}~\bibnamefont
  {Watanabe}},\ and\ \bibinfo {author} {\bibfnamefont {A.~F.}\ \bibnamefont
  {Young}},\ }\bibfield  {title} {\bibinfo {title} {Isospin magnetism and
  spin-polarized superconductivity in bernal bilayer graphene},\ }\href
  {https://doi.org/10.1126/science.abm8386} {\bibfield  {journal} {\bibinfo
  {journal} {Science}\ }\textbf {\bibinfo {volume} {375}},\ \bibinfo {pages}
  {774} (\bibinfo {year} {2022})},\ \Eprint
  {https://arxiv.org/abs/https://www.science.org/doi/pdf/10.1126/science.abm8386}
  {https://www.science.org/doi/pdf/10.1126/science.abm8386} \BibitemShut
  {NoStop}%
\bibitem [{\citenamefont {{Zhang}}\ \emph {et~al.}(2022)\citenamefont
  {{Zhang}}, \citenamefont {{Polski}}, \citenamefont {{Thomson}}, \citenamefont
  {{Lantagne-Hurtubise}}, \citenamefont {{Lewandowski}}, \citenamefont
  {{Zhou}}, \citenamefont {{Watanabe}}, \citenamefont {{Taniguchi}},
  \citenamefont {{Alicea}},\ and\ \citenamefont {{Nadj-Perge}}}]{Zhang_cm22}%
  \BibitemOpen
  \bibfield  {author} {\bibinfo {author} {\bibfnamefont {Y.}~\bibnamefont
  {{Zhang}}}, \bibinfo {author} {\bibfnamefont {R.}~\bibnamefont {{Polski}}},
  \bibinfo {author} {\bibfnamefont {A.}~\bibnamefont {{Thomson}}}, \bibinfo
  {author} {\bibfnamefont {{\'E}.}~\bibnamefont {{Lantagne-Hurtubise}}},
  \bibinfo {author} {\bibfnamefont {C.}~\bibnamefont {{Lewandowski}}}, \bibinfo
  {author} {\bibfnamefont {H.}~\bibnamefont {{Zhou}}}, \bibinfo {author}
  {\bibfnamefont {K.}~\bibnamefont {{Watanabe}}}, \bibinfo {author}
  {\bibfnamefont {T.}~\bibnamefont {{Taniguchi}}}, \bibinfo {author}
  {\bibfnamefont {J.}~\bibnamefont {{Alicea}}},\ and\ \bibinfo {author}
  {\bibfnamefont {S.}~\bibnamefont {{Nadj-Perge}}},\ }\bibfield  {title}
  {\bibinfo {title} {{Spin-Orbit Enhanced Superconductivity in Bernal Bilayer
  Graphene}},\ }\href@noop {} {\bibfield  {journal} {\bibinfo  {journal} {arXiv
  e-prints}\ ,\ \bibinfo {eid} {arXiv:2205.05087}} (\bibinfo {year} {2022})},\
  \Eprint {https://arxiv.org/abs/2205.05087} {arXiv:2205.05087
  [cond-mat.supr-con]} \BibitemShut {NoStop}%
\bibitem [{\citenamefont {Guinea}\ and\ \citenamefont
  {Walet}(2018)}]{Guinea_pnas18}%
  \BibitemOpen
  \bibfield  {author} {\bibinfo {author} {\bibfnamefont {F.}~\bibnamefont
  {Guinea}}\ and\ \bibinfo {author} {\bibfnamefont {N.~R.}\ \bibnamefont
  {Walet}},\ }\bibfield  {title} {\bibinfo {title} {Electrostatic effects, band
  distortions, and superconductivity in twisted graphene bilayers},\ }\href
  {https://doi.org/10.1073/pnas.1810947115} {\bibfield  {journal} {\bibinfo
  {journal} {Proceedings of the National Academy of Sciences}\ }\textbf
  {\bibinfo {volume} {115}},\ \bibinfo {pages} {13174} (\bibinfo {year}
  {2018})},\ \Eprint
  {https://arxiv.org/abs/https://www.pnas.org/doi/pdf/10.1073/pnas.1810947115}
  {https://www.pnas.org/doi/pdf/10.1073/pnas.1810947115} \BibitemShut {NoStop}%
\bibitem [{\citenamefont {Sherkunov}\ and\ \citenamefont
  {Betouras}(2018)}]{Sherkunov_prb18}%
  \BibitemOpen
  \bibfield  {author} {\bibinfo {author} {\bibfnamefont {Y.}~\bibnamefont
  {Sherkunov}}\ and\ \bibinfo {author} {\bibfnamefont {J.~J.}\ \bibnamefont
  {Betouras}},\ }\bibfield  {title} {\bibinfo {title} {Electronic phases in
  twisted bilayer graphene at magic angles as a result of van hove
  singularities and interactions},\ }\href
  {https://doi.org/10.1103/PhysRevB.98.205151} {\bibfield  {journal} {\bibinfo
  {journal} {Phys. Rev. B}\ }\textbf {\bibinfo {volume} {98}},\ \bibinfo
  {pages} {205151} (\bibinfo {year} {2018})}\BibitemShut {NoStop}%
\bibitem [{\citenamefont {Cea}\ \emph {et~al.}(2019)\citenamefont {Cea},
  \citenamefont {Walet},\ and\ \citenamefont {Guinea}}]{Cea_prb19}%
  \BibitemOpen
  \bibfield  {author} {\bibinfo {author} {\bibfnamefont {T.}~\bibnamefont
  {Cea}}, \bibinfo {author} {\bibfnamefont {N.~R.}\ \bibnamefont {Walet}},\
  and\ \bibinfo {author} {\bibfnamefont {F.}~\bibnamefont {Guinea}},\
  }\bibfield  {title} {\bibinfo {title} {Electronic band structure and pinning
  of fermi energy to van hove singularities in twisted bilayer graphene: A
  self-consistent approach},\ }\href
  {https://doi.org/10.1103/PhysRevB.100.205113} {\bibfield  {journal} {\bibinfo
   {journal} {Phys. Rev. B}\ }\textbf {\bibinfo {volume} {100}},\ \bibinfo
  {pages} {205113} (\bibinfo {year} {2019})}\BibitemShut {NoStop}%
\bibitem [{\citenamefont {Cea}\ and\ \citenamefont {Guinea}(2020)}]{Cea_prb20}%
  \BibitemOpen
  \bibfield  {author} {\bibinfo {author} {\bibfnamefont {T.}~\bibnamefont
  {Cea}}\ and\ \bibinfo {author} {\bibfnamefont {F.}~\bibnamefont {Guinea}},\
  }\bibfield  {title} {\bibinfo {title} {Band structure and insulating states
  driven by coulomb interaction in twisted bilayer graphene},\ }\href
  {https://doi.org/10.1103/PhysRevB.102.045107} {\bibfield  {journal} {\bibinfo
   {journal} {Phys. Rev. B}\ }\textbf {\bibinfo {volume} {102}},\ \bibinfo
  {pages} {045107} (\bibinfo {year} {2020})}\BibitemShut {NoStop}%
\bibitem [{\citenamefont {Lin}\ and\ \citenamefont
  {Nandkishore}(2020)}]{Lin_prb20}%
  \BibitemOpen
  \bibfield  {author} {\bibinfo {author} {\bibfnamefont {Y.-P.}\ \bibnamefont
  {Lin}}\ and\ \bibinfo {author} {\bibfnamefont {R.~M.}\ \bibnamefont
  {Nandkishore}},\ }\bibfield  {title} {\bibinfo {title} {Parquet
  renormalization group analysis of weak-coupling instabilities with multiple
  high-order van hove points inside the brillouin zone},\ }\href
  {https://doi.org/10.1103/PhysRevB.102.245122} {\bibfield  {journal} {\bibinfo
   {journal} {Phys. Rev. B}\ }\textbf {\bibinfo {volume} {102}},\ \bibinfo
  {pages} {245122} (\bibinfo {year} {2020})}\BibitemShut {NoStop}%
\bibitem [{\citenamefont {{Chichinadze}}\ \emph {et~al.}(2022)\citenamefont
  {{Chichinadze}}, \citenamefont {{Classen}}, \citenamefont {{Wang}},\ and\
  \citenamefont {{Chubukov}}}]{Chichinadze_cm22}%
  \BibitemOpen
  \bibfield  {author} {\bibinfo {author} {\bibfnamefont {D.~V.}\ \bibnamefont
  {{Chichinadze}}}, \bibinfo {author} {\bibfnamefont {L.}~\bibnamefont
  {{Classen}}}, \bibinfo {author} {\bibfnamefont {Y.}~\bibnamefont {{Wang}}},\
  and\ \bibinfo {author} {\bibfnamefont {A.~V.}\ \bibnamefont {{Chubukov}}},\
  }\bibfield  {title} {\bibinfo {title} {{Cascade of transitions in twisted
  bilayer graphene -- the Van Hove scenario}},\ }\href@noop {} {\bibfield
  {journal} {\bibinfo  {journal} {arXiv e-prints}\ ,\ \bibinfo {eid}
  {arXiv:2206.10539}} (\bibinfo {year} {2022})},\ \Eprint
  {https://arxiv.org/abs/2206.10539} {arXiv:2206.10539 [cond-mat.mes-hall]}
  \BibitemShut {NoStop}%
\bibitem [{\citenamefont {Isobe}\ \emph {et~al.}(2018)\citenamefont {Isobe},
  \citenamefont {Yuan},\ and\ \citenamefont {Fu}}]{Isobe_prx18}%
  \BibitemOpen
  \bibfield  {author} {\bibinfo {author} {\bibfnamefont {H.}~\bibnamefont
  {Isobe}}, \bibinfo {author} {\bibfnamefont {N.~F.~Q.}\ \bibnamefont {Yuan}},\
  and\ \bibinfo {author} {\bibfnamefont {L.}~\bibnamefont {Fu}},\ }\bibfield
  {title} {\bibinfo {title} {Unconventional superconductivity and density waves
  in twisted bilayer graphene},\ }\href
  {https://doi.org/10.1103/PhysRevX.8.041041} {\bibfield  {journal} {\bibinfo
  {journal} {Phys. Rev. X}\ }\textbf {\bibinfo {volume} {8}},\ \bibinfo {pages}
  {041041} (\bibinfo {year} {2018})}\BibitemShut {NoStop}%
\bibitem [{\citenamefont {Liu}\ \emph {et~al.}(2018)\citenamefont {Liu},
  \citenamefont {Zhang}, \citenamefont {Chen},\ and\ \citenamefont
  {Yang}}]{Liu_prl18}%
  \BibitemOpen
  \bibfield  {author} {\bibinfo {author} {\bibfnamefont {C.-C.}\ \bibnamefont
  {Liu}}, \bibinfo {author} {\bibfnamefont {L.-D.}\ \bibnamefont {Zhang}},
  \bibinfo {author} {\bibfnamefont {W.-Q.}\ \bibnamefont {Chen}},\ and\
  \bibinfo {author} {\bibfnamefont {F.}~\bibnamefont {Yang}},\ }\bibfield
  {title} {\bibinfo {title} {Chiral spin density wave and $d+id$
  superconductivity in the magic-angle-twisted bilayer graphene},\ }\href
  {https://doi.org/10.1103/PhysRevLett.121.217001} {\bibfield  {journal}
  {\bibinfo  {journal} {Phys. Rev. Lett.}\ }\textbf {\bibinfo {volume} {121}},\
  \bibinfo {pages} {217001} (\bibinfo {year} {2018})}\BibitemShut {NoStop}%
\bibitem [{\citenamefont {Gonz\'alez}\ and\ \citenamefont
  {Stauber}(2019)}]{Gonzalez_prl19}%
  \BibitemOpen
  \bibfield  {author} {\bibinfo {author} {\bibfnamefont {J.}~\bibnamefont
  {Gonz\'alez}}\ and\ \bibinfo {author} {\bibfnamefont {T.}~\bibnamefont
  {Stauber}},\ }\bibfield  {title} {\bibinfo {title} {Kohn-luttinger
  superconductivity in twisted bilayer graphene},\ }\href
  {https://doi.org/10.1103/PhysRevLett.122.026801} {\bibfield  {journal}
  {\bibinfo  {journal} {Phys. Rev. Lett.}\ }\textbf {\bibinfo {volume} {122}},\
  \bibinfo {pages} {026801} (\bibinfo {year} {2019})}\BibitemShut {NoStop}%
\bibitem [{\citenamefont {You}\ and\ \citenamefont
  {Vishwanath}(2019)}]{You_npj19}%
  \BibitemOpen
  \bibfield  {author} {\bibinfo {author} {\bibfnamefont {Y.-Z.}\ \bibnamefont
  {You}}\ and\ \bibinfo {author} {\bibfnamefont {A.}~\bibnamefont
  {Vishwanath}},\ }\bibfield  {title} {\bibinfo {title} {Superconductivity from
  valley fluctuations and approximate so(4) symmetry in a weak coupling theory
  of twisted bilayer graphene},\ }\href
  {https://doi.org/10.1038/s41535-019-0153-4} {\bibfield  {journal} {\bibinfo
  {journal} {npj Quantum Materials}\ }\textbf {\bibinfo {volume} {4}},\
  \bibinfo {pages} {16} (\bibinfo {year} {2019})}\BibitemShut {NoStop}%
\bibitem [{\citenamefont {Roy}\ and\ \citenamefont {Juri\ifmmode \check{c}\else
  \v{c}\fi{}i\ifmmode~\acute{c}\else \'{c}\fi{}}(2019)}]{Roy_prb19}%
  \BibitemOpen
  \bibfield  {author} {\bibinfo {author} {\bibfnamefont {B.}~\bibnamefont
  {Roy}}\ and\ \bibinfo {author} {\bibfnamefont {V.}~\bibnamefont {Juri\ifmmode
  \check{c}\else \v{c}\fi{}i\ifmmode~\acute{c}\else \'{c}\fi{}}},\ }\bibfield
  {title} {\bibinfo {title} {Unconventional superconductivity in nearly flat
  bands in twisted bilayer graphene},\ }\href
  {https://doi.org/10.1103/PhysRevB.99.121407} {\bibfield  {journal} {\bibinfo
  {journal} {Phys. Rev. B}\ }\textbf {\bibinfo {volume} {99}},\ \bibinfo
  {pages} {121407} (\bibinfo {year} {2019})}\BibitemShut {NoStop}%
\bibitem [{\citenamefont {Sharma}\ \emph {et~al.}(2020)\citenamefont {Sharma},
  \citenamefont {Trushin}, \citenamefont {Sushkov}, \citenamefont {Vignale},\
  and\ \citenamefont {Adam}}]{Sharma_prr20}%
  \BibitemOpen
  \bibfield  {author} {\bibinfo {author} {\bibfnamefont {G.}~\bibnamefont
  {Sharma}}, \bibinfo {author} {\bibfnamefont {M.}~\bibnamefont {Trushin}},
  \bibinfo {author} {\bibfnamefont {O.~P.}\ \bibnamefont {Sushkov}}, \bibinfo
  {author} {\bibfnamefont {G.}~\bibnamefont {Vignale}},\ and\ \bibinfo {author}
  {\bibfnamefont {S.}~\bibnamefont {Adam}},\ }\bibfield  {title} {\bibinfo
  {title} {Superconductivity from collective excitations in magic-angle twisted
  bilayer graphene},\ }\href {https://doi.org/10.1103/PhysRevResearch.2.022040}
  {\bibfield  {journal} {\bibinfo  {journal} {Phys. Rev. Research}\ }\textbf
  {\bibinfo {volume} {2}},\ \bibinfo {pages} {022040} (\bibinfo {year}
  {2020})}\BibitemShut {NoStop}%
\bibitem [{\citenamefont {Chichinadze}\ \emph {et~al.}(2020)\citenamefont
  {Chichinadze}, \citenamefont {Classen},\ and\ \citenamefont
  {Chubukov}}]{Chichinadze_prb20}%
  \BibitemOpen
  \bibfield  {author} {\bibinfo {author} {\bibfnamefont {D.~V.}\ \bibnamefont
  {Chichinadze}}, \bibinfo {author} {\bibfnamefont {L.}~\bibnamefont
  {Classen}},\ and\ \bibinfo {author} {\bibfnamefont {A.~V.}\ \bibnamefont
  {Chubukov}},\ }\bibfield  {title} {\bibinfo {title} {Nematic
  superconductivity in twisted bilayer graphene},\ }\href
  {https://doi.org/10.1103/PhysRevB.101.224513} {\bibfield  {journal} {\bibinfo
   {journal} {Phys. Rev. B}\ }\textbf {\bibinfo {volume} {101}},\ \bibinfo
  {pages} {224513} (\bibinfo {year} {2020})}\BibitemShut {NoStop}%
\bibitem [{\citenamefont {{Qin}}\ \emph {et~al.}(2021)\citenamefont {{Qin}},
  \citenamefont {{Zou}},\ and\ \citenamefont {{MacDonald}}}]{Qin_cm21}%
  \BibitemOpen
  \bibfield  {author} {\bibinfo {author} {\bibfnamefont {W.}~\bibnamefont
  {{Qin}}}, \bibinfo {author} {\bibfnamefont {B.}~\bibnamefont {{Zou}}},\ and\
  \bibinfo {author} {\bibfnamefont {A.~H.}\ \bibnamefont {{MacDonald}}},\
  }\bibfield  {title} {\bibinfo {title} {{Critical magnetic fields and
  electron-pairing in magic-angle twisted bilayer graphene}},\ }\href@noop {}
  {\bibfield  {journal} {\bibinfo  {journal} {arXiv e-prints}\ ,\ \bibinfo
  {eid} {arXiv:2102.10504}} (\bibinfo {year} {2021})},\ \Eprint
  {https://arxiv.org/abs/2102.10504} {arXiv:2102.10504 [cond-mat.supr-con]}
  \BibitemShut {NoStop}%
\bibitem [{\citenamefont {{Dong}}\ and\ \citenamefont
  {{Levitov}}(2021{\natexlab{a}})}]{Dong_cm21}%
  \BibitemOpen
  \bibfield  {author} {\bibinfo {author} {\bibfnamefont {Z.}~\bibnamefont
  {{Dong}}}\ and\ \bibinfo {author} {\bibfnamefont {L.}~\bibnamefont
  {{Levitov}}},\ }\bibfield  {title} {\bibinfo {title} {{Activating
  superconductivity in a repulsive system by high-energy degrees of freedom}},\
  }\href@noop {} {\bibfield  {journal} {\bibinfo  {journal} {arXiv e-prints}\
  ,\ \bibinfo {eid} {arXiv:2103.08767}} (\bibinfo {year}
  {2021}{\natexlab{a}})},\ \Eprint {https://arxiv.org/abs/2103.08767}
  {arXiv:2103.08767 [cond-mat.supr-con]} \BibitemShut {NoStop}%
\bibitem [{\citenamefont {Ghazaryan}\ \emph {et~al.}(2021)\citenamefont
  {Ghazaryan}, \citenamefont {Holder}, \citenamefont {Serbyn},\ and\
  \citenamefont {Berg}}]{Ghazaryan_prl21}%
  \BibitemOpen
  \bibfield  {author} {\bibinfo {author} {\bibfnamefont {A.}~\bibnamefont
  {Ghazaryan}}, \bibinfo {author} {\bibfnamefont {T.}~\bibnamefont {Holder}},
  \bibinfo {author} {\bibfnamefont {M.}~\bibnamefont {Serbyn}},\ and\ \bibinfo
  {author} {\bibfnamefont {E.}~\bibnamefont {Berg}},\ }\bibfield  {title}
  {\bibinfo {title} {Unconventional superconductivity in systems with annular
  fermi surfaces: Application to rhombohedral trilayer graphene},\ }\href
  {https://doi.org/10.1103/PhysRevLett.127.247001} {\bibfield  {journal}
  {\bibinfo  {journal} {Phys. Rev. Lett.}\ }\textbf {\bibinfo {volume} {127}},\
  \bibinfo {pages} {247001} (\bibinfo {year} {2021})}\BibitemShut {NoStop}%
\bibitem [{\citenamefont {Dai}\ \emph {et~al.}(2021)\citenamefont {Dai},
  \citenamefont {Hou}, \citenamefont {Zhang}, \citenamefont {Liang},\ and\
  \citenamefont {Ma}}]{Dai_prl21}%
  \BibitemOpen
  \bibfield  {author} {\bibinfo {author} {\bibfnamefont {H.}~\bibnamefont
  {Dai}}, \bibinfo {author} {\bibfnamefont {J.}~\bibnamefont {Hou}}, \bibinfo
  {author} {\bibfnamefont {X.}~\bibnamefont {Zhang}}, \bibinfo {author}
  {\bibfnamefont {Y.}~\bibnamefont {Liang}},\ and\ \bibinfo {author}
  {\bibfnamefont {T.}~\bibnamefont {Ma}},\ }\bibfield  {title} {\bibinfo
  {title} {Mott insulating state and $d+id$ superconductivity in an abc
  graphene trilayer},\ }\href {https://doi.org/10.1103/PhysRevB.104.035104}
  {\bibfield  {journal} {\bibinfo  {journal} {Phys. Rev. B}\ }\textbf {\bibinfo
  {volume} {104}},\ \bibinfo {pages} {035104} (\bibinfo {year}
  {2021})}\BibitemShut {NoStop}%
\bibitem [{\citenamefont {{Gonzalez}}\ and\ \citenamefont
  {{Stauber}}(2021)}]{Gonzalez_cm21}%
  \BibitemOpen
  \bibfield  {author} {\bibinfo {author} {\bibfnamefont {J.}~\bibnamefont
  {{Gonzalez}}}\ and\ \bibinfo {author} {\bibfnamefont {T.}~\bibnamefont
  {{Stauber}}},\ }\bibfield  {title} {\bibinfo {title} {{Ising
  superconductivity induced from valley symmetry breaking in twisted trilayer
  graphene}},\ }\href@noop {} {\bibfield  {journal} {\bibinfo  {journal} {arXiv
  e-prints}\ ,\ \bibinfo {eid} {arXiv:2110.11294}} (\bibinfo {year} {2021})},\
  \Eprint {https://arxiv.org/abs/2110.11294} {arXiv:2110.11294
  [cond-mat.supr-con]} \BibitemShut {NoStop}%
\bibitem [{\citenamefont {Fischer}\ \emph {et~al.}(2022)\citenamefont
  {Fischer}, \citenamefont {Goodwin}, \citenamefont {Mostofi}, \citenamefont
  {Lischner}, \citenamefont {Kennes},\ and\ \citenamefont
  {Klebl}}]{Fischer_npj22}%
  \BibitemOpen
  \bibfield  {author} {\bibinfo {author} {\bibfnamefont {A.}~\bibnamefont
  {Fischer}}, \bibinfo {author} {\bibfnamefont {Z.~A.~H.}\ \bibnamefont
  {Goodwin}}, \bibinfo {author} {\bibfnamefont {A.~A.}\ \bibnamefont
  {Mostofi}}, \bibinfo {author} {\bibfnamefont {J.}~\bibnamefont {Lischner}},
  \bibinfo {author} {\bibfnamefont {D.~M.}\ \bibnamefont {Kennes}},\ and\
  \bibinfo {author} {\bibfnamefont {L.}~\bibnamefont {Klebl}},\ }\bibfield
  {title} {\bibinfo {title} {Unconventional superconductivity in magic-angle
  twisted trilayer graphene},\ }\href
  {https://doi.org/10.1038/s41535-021-00410-w} {\bibfield  {journal} {\bibinfo
  {journal} {npj Quantum Materials}\ }\textbf {\bibinfo {volume} {7}},\
  \bibinfo {pages} {5} (\bibinfo {year} {2022})}\BibitemShut {NoStop}%
\bibitem [{\citenamefont {Szab\'o}\ and\ \citenamefont
  {Roy}(2022)}]{Szabo_prb22}%
  \BibitemOpen
  \bibfield  {author} {\bibinfo {author} {\bibfnamefont {A.~L.}\ \bibnamefont
  {Szab\'o}}\ and\ \bibinfo {author} {\bibfnamefont {B.}~\bibnamefont {Roy}},\
  }\bibfield  {title} {\bibinfo {title} {Metals, fractional metals, and
  superconductivity in rhombohedral trilayer graphene},\ }\href
  {https://doi.org/10.1103/PhysRevB.105.L081407} {\bibfield  {journal}
  {\bibinfo  {journal} {Phys. Rev. B}\ }\textbf {\bibinfo {volume} {105}},\
  \bibinfo {pages} {L081407} (\bibinfo {year} {2022})}\BibitemShut {NoStop}%
\bibitem [{\citenamefont {Cea}\ \emph {et~al.}(2022)\citenamefont {Cea},
  \citenamefont {Pantale\'on}, \citenamefont {Phong},\ and\ \citenamefont
  {Guinea}}]{Cea_prb22}%
  \BibitemOpen
  \bibfield  {author} {\bibinfo {author} {\bibfnamefont {T.}~\bibnamefont
  {Cea}}, \bibinfo {author} {\bibfnamefont {P.~A.}\ \bibnamefont
  {Pantale\'on}}, \bibinfo {author} {\bibfnamefont {V.~o.~T.}\ \bibnamefont
  {Phong}},\ and\ \bibinfo {author} {\bibfnamefont {F.}~\bibnamefont
  {Guinea}},\ }\bibfield  {title} {\bibinfo {title} {Superconductivity from
  repulsive interactions in rhombohedral trilayer graphene: A
  kohn-luttinger-like mechanism},\ }\href
  {https://doi.org/10.1103/PhysRevB.105.075432} {\bibfield  {journal} {\bibinfo
   {journal} {Phys. Rev. B}\ }\textbf {\bibinfo {volume} {105}},\ \bibinfo
  {pages} {075432} (\bibinfo {year} {2022})}\BibitemShut {NoStop}%
\bibitem [{\citenamefont {{Qin}}\ \emph {et~al.}(2022)\citenamefont {{Qin}},
  \citenamefont {{Huang}}, \citenamefont {{Wolf}}, \citenamefont {{Wei}},
  \citenamefont {{Blinov}},\ and\ \citenamefont {{MacDonald}}}]{Qin_cm22}%
  \BibitemOpen
  \bibfield  {author} {\bibinfo {author} {\bibfnamefont {W.}~\bibnamefont
  {{Qin}}}, \bibinfo {author} {\bibfnamefont {C.}~\bibnamefont {{Huang}}},
  \bibinfo {author} {\bibfnamefont {T.}~\bibnamefont {{Wolf}}}, \bibinfo
  {author} {\bibfnamefont {N.}~\bibnamefont {{Wei}}}, \bibinfo {author}
  {\bibfnamefont {I.}~\bibnamefont {{Blinov}}},\ and\ \bibinfo {author}
  {\bibfnamefont {A.~H.}\ \bibnamefont {{MacDonald}}},\ }\bibfield  {title}
  {\bibinfo {title} {{Functional Renormalization Group Study of
  Superconductivity in Rhombohedral Trilayer Graphene}},\ }\href@noop {}
  {\bibfield  {journal} {\bibinfo  {journal} {arXiv e-prints}\ ,\ \bibinfo
  {eid} {arXiv:2203.09083}} (\bibinfo {year} {2022})},\ \Eprint
  {https://arxiv.org/abs/2203.09083} {arXiv:2203.09083 [cond-mat.supr-con]}
  \BibitemShut {NoStop}%
\bibitem [{\citenamefont {Cr\'epel}\ \emph {et~al.}(2022)\citenamefont
  {Cr\'epel}, \citenamefont {Cea}, \citenamefont {Fu},\ and\ \citenamefont
  {Guinea}}]{Crepel_prb22}%
  \BibitemOpen
  \bibfield  {author} {\bibinfo {author} {\bibfnamefont {V.}~\bibnamefont
  {Cr\'epel}}, \bibinfo {author} {\bibfnamefont {T.}~\bibnamefont {Cea}},
  \bibinfo {author} {\bibfnamefont {L.}~\bibnamefont {Fu}},\ and\ \bibinfo
  {author} {\bibfnamefont {F.}~\bibnamefont {Guinea}},\ }\bibfield  {title}
  {\bibinfo {title} {Unconventional superconductivity due to interband
  polarization},\ }\href {https://doi.org/10.1103/PhysRevB.105.094506}
  {\bibfield  {journal} {\bibinfo  {journal} {Phys. Rev. B}\ }\textbf {\bibinfo
  {volume} {105}},\ \bibinfo {pages} {094506} (\bibinfo {year}
  {2022})}\BibitemShut {NoStop}%
\bibitem [{\citenamefont {Lu}\ \emph {et~al.}(2022)\citenamefont {Lu},
  \citenamefont {Wang}, \citenamefont {Chatterjee},\ and\ \citenamefont
  {You}}]{Lu_prb22}%
  \BibitemOpen
  \bibfield  {author} {\bibinfo {author} {\bibfnamefont {D.-C.}\ \bibnamefont
  {Lu}}, \bibinfo {author} {\bibfnamefont {T.}~\bibnamefont {Wang}}, \bibinfo
  {author} {\bibfnamefont {S.}~\bibnamefont {Chatterjee}},\ and\ \bibinfo
  {author} {\bibfnamefont {Y.-Z.}\ \bibnamefont {You}},\ }\bibfield  {title}
  {\bibinfo {title} {Correlated metals and unconventional superconductivity in
  rhombohedral trilayer graphene: A renormalization group analysis},\ }\href
  {https://doi.org/10.1103/PhysRevB.106.155115} {\bibfield  {journal} {\bibinfo
   {journal} {Phys. Rev. B}\ }\textbf {\bibinfo {volume} {106}},\ \bibinfo
  {pages} {155115} (\bibinfo {year} {2022})}\BibitemShut {NoStop}%
\bibitem [{\citenamefont {{Jimeno-Pozo}}\ \emph {et~al.}(2022)\citenamefont
  {{Jimeno-Pozo}}, \citenamefont {{Sainz-Cruz}}, \citenamefont {{Cea}},
  \citenamefont {{Pantale{\'o}n}},\ and\ \citenamefont
  {{Guinea}}}]{Jimeno_cm22}%
  \BibitemOpen
  \bibfield  {author} {\bibinfo {author} {\bibfnamefont {A.}~\bibnamefont
  {{Jimeno-Pozo}}}, \bibinfo {author} {\bibfnamefont {H.}~\bibnamefont
  {{Sainz-Cruz}}}, \bibinfo {author} {\bibfnamefont {T.}~\bibnamefont {{Cea}}},
  \bibinfo {author} {\bibfnamefont {P.~A.}\ \bibnamefont {{Pantale{\'o}n}}},\
  and\ \bibinfo {author} {\bibfnamefont {F.}~\bibnamefont {{Guinea}}},\
  }\bibfield  {title} {\bibinfo {title} {{Superconductivity from electronic
  interactions and spin-orbit enhancement in bilayer and trilayer graphene}},\
  }\href@noop {} {\bibfield  {journal} {\bibinfo  {journal} {arXiv e-prints}\
  ,\ \bibinfo {eid} {arXiv:2210.02915}} (\bibinfo {year} {2022})},\ \Eprint
  {https://arxiv.org/abs/2210.02915} {arXiv:2210.02915 [cond-mat.mes-hall]}
  \BibitemShut {NoStop}%
\bibitem [{\citenamefont {Wu}\ \emph {et~al.}(2018)\citenamefont {Wu},
  \citenamefont {MacDonald},\ and\ \citenamefont {Martin}}]{Wu_prl18}%
  \BibitemOpen
  \bibfield  {author} {\bibinfo {author} {\bibfnamefont {F.}~\bibnamefont
  {Wu}}, \bibinfo {author} {\bibfnamefont {A.~H.}\ \bibnamefont {MacDonald}},\
  and\ \bibinfo {author} {\bibfnamefont {I.}~\bibnamefont {Martin}},\
  }\bibfield  {title} {\bibinfo {title} {Theory of phonon-mediated
  superconductivity in twisted bilayer graphene},\ }\href
  {https://doi.org/10.1103/PhysRevLett.121.257001} {\bibfield  {journal}
  {\bibinfo  {journal} {Phys. Rev. Lett.}\ }\textbf {\bibinfo {volume} {121}},\
  \bibinfo {pages} {257001} (\bibinfo {year} {2018})}\BibitemShut {NoStop}%
\bibitem [{\citenamefont {Peltonen}\ \emph {et~al.}(2018)\citenamefont
  {Peltonen}, \citenamefont {Ojaj\"arvi},\ and\ \citenamefont
  {Heikkil\"a}}]{Peltonen_prb18}%
  \BibitemOpen
  \bibfield  {author} {\bibinfo {author} {\bibfnamefont {T.~J.}\ \bibnamefont
  {Peltonen}}, \bibinfo {author} {\bibfnamefont {R.}~\bibnamefont
  {Ojaj\"arvi}},\ and\ \bibinfo {author} {\bibfnamefont {T.~T.}\ \bibnamefont
  {Heikkil\"a}},\ }\bibfield  {title} {\bibinfo {title} {Mean-field theory for
  superconductivity in twisted bilayer graphene},\ }\href
  {https://doi.org/10.1103/PhysRevB.98.220504} {\bibfield  {journal} {\bibinfo
  {journal} {Phys. Rev. B}\ }\textbf {\bibinfo {volume} {98}},\ \bibinfo
  {pages} {220504} (\bibinfo {year} {2018})}\BibitemShut {NoStop}%
\bibitem [{\citenamefont {Choi}\ and\ \citenamefont {Choi}(2018)}]{Choi_prb18}%
  \BibitemOpen
  \bibfield  {author} {\bibinfo {author} {\bibfnamefont {Y.~W.}\ \bibnamefont
  {Choi}}\ and\ \bibinfo {author} {\bibfnamefont {H.~J.}\ \bibnamefont
  {Choi}},\ }\bibfield  {title} {\bibinfo {title} {Strong electron-phonon
  coupling, electron-hole asymmetry, and nonadiabaticity in magic-angle twisted
  bilayer graphene},\ }\href {https://doi.org/10.1103/PhysRevB.98.241412}
  {\bibfield  {journal} {\bibinfo  {journal} {Phys. Rev. B}\ }\textbf {\bibinfo
  {volume} {98}},\ \bibinfo {pages} {241412} (\bibinfo {year}
  {2018})}\BibitemShut {NoStop}%
\bibitem [{\citenamefont {Lian}\ \emph {et~al.}(2019)\citenamefont {Lian},
  \citenamefont {Wang},\ and\ \citenamefont {Bernevig}}]{Lian_prl19}%
  \BibitemOpen
  \bibfield  {author} {\bibinfo {author} {\bibfnamefont {B.}~\bibnamefont
  {Lian}}, \bibinfo {author} {\bibfnamefont {Z.}~\bibnamefont {Wang}},\ and\
  \bibinfo {author} {\bibfnamefont {B.~A.}\ \bibnamefont {Bernevig}},\
  }\bibfield  {title} {\bibinfo {title} {Twisted bilayer graphene: A
  phonon-driven superconductor},\ }\href
  {https://doi.org/10.1103/PhysRevLett.122.257002} {\bibfield  {journal}
  {\bibinfo  {journal} {Phys. Rev. Lett.}\ }\textbf {\bibinfo {volume} {122}},\
  \bibinfo {pages} {257002} (\bibinfo {year} {2019})}\BibitemShut {NoStop}%
\bibitem [{\citenamefont {Angeli}\ \emph {et~al.}(2019)\citenamefont {Angeli},
  \citenamefont {Tosatti},\ and\ \citenamefont {Fabrizio}}]{Angeli_prx19}%
  \BibitemOpen
  \bibfield  {author} {\bibinfo {author} {\bibfnamefont {M.}~\bibnamefont
  {Angeli}}, \bibinfo {author} {\bibfnamefont {E.}~\bibnamefont {Tosatti}},\
  and\ \bibinfo {author} {\bibfnamefont {M.}~\bibnamefont {Fabrizio}},\
  }\bibfield  {title} {\bibinfo {title} {Valley jahn-teller effect in twisted
  bilayer graphene},\ }\href {https://doi.org/10.1103/PhysRevX.9.041010}
  {\bibfield  {journal} {\bibinfo  {journal} {Phys. Rev. X}\ }\textbf {\bibinfo
  {volume} {9}},\ \bibinfo {pages} {041010} (\bibinfo {year}
  {2019})}\BibitemShut {NoStop}%
\bibitem [{\citenamefont {Wu}\ \emph {et~al.}(2019)\citenamefont {Wu},
  \citenamefont {Hwang},\ and\ \citenamefont {Das~Sarma}}]{Wu_prb19}%
  \BibitemOpen
  \bibfield  {author} {\bibinfo {author} {\bibfnamefont {F.}~\bibnamefont
  {Wu}}, \bibinfo {author} {\bibfnamefont {E.}~\bibnamefont {Hwang}},\ and\
  \bibinfo {author} {\bibfnamefont {S.}~\bibnamefont {Das~Sarma}},\ }\bibfield
  {title} {\bibinfo {title} {Phonon-induced giant linear-in-$t$ resistivity in
  magic angle twisted bilayer graphene: Ordinary strangeness and exotic
  superconductivity},\ }\href {https://doi.org/10.1103/PhysRevB.99.165112}
  {\bibfield  {journal} {\bibinfo  {journal} {Phys. Rev. B}\ }\textbf {\bibinfo
  {volume} {99}},\ \bibinfo {pages} {165112} (\bibinfo {year}
  {2019})}\BibitemShut {NoStop}%
\bibitem [{\citenamefont {Schrodi}\ \emph {et~al.}(2020)\citenamefont
  {Schrodi}, \citenamefont {Aperis},\ and\ \citenamefont
  {Oppeneer}}]{Schrodi_prr20}%
  \BibitemOpen
  \bibfield  {author} {\bibinfo {author} {\bibfnamefont {F.}~\bibnamefont
  {Schrodi}}, \bibinfo {author} {\bibfnamefont {A.}~\bibnamefont {Aperis}},\
  and\ \bibinfo {author} {\bibfnamefont {P.~M.}\ \bibnamefont {Oppeneer}},\
  }\bibfield  {title} {\bibinfo {title} {Prominent cooper pairing away from the
  fermi level and its spectroscopic signature in twisted bilayer graphene},\
  }\href {https://doi.org/10.1103/PhysRevResearch.2.012066} {\bibfield
  {journal} {\bibinfo  {journal} {Phys. Rev. Research}\ }\textbf {\bibinfo
  {volume} {2}},\ \bibinfo {pages} {012066} (\bibinfo {year}
  {2020})}\BibitemShut {NoStop}%
\bibitem [{\citenamefont {Choi}\ and\ \citenamefont {Choi}(2021)}]{Choi_prl21}%
  \BibitemOpen
  \bibfield  {author} {\bibinfo {author} {\bibfnamefont {Y.~W.}\ \bibnamefont
  {Choi}}\ and\ \bibinfo {author} {\bibfnamefont {H.~J.}\ \bibnamefont
  {Choi}},\ }\bibfield  {title} {\bibinfo {title} {Dichotomy of electron-phonon
  coupling in graphene moir\'e flat bands},\ }\href
  {https://doi.org/10.1103/PhysRevLett.127.167001} {\bibfield  {journal}
  {\bibinfo  {journal} {Phys. Rev. Lett.}\ }\textbf {\bibinfo {volume} {127}},\
  \bibinfo {pages} {167001} (\bibinfo {year} {2021})}\BibitemShut {NoStop}%
\bibitem [{\citenamefont {Chou}\ \emph {et~al.}(2021)\citenamefont {Chou},
  \citenamefont {Wu}, \citenamefont {Sau},\ and\ \citenamefont
  {Das~Sarma}}]{Chou_prl21}%
  \BibitemOpen
  \bibfield  {author} {\bibinfo {author} {\bibfnamefont {Y.-Z.}\ \bibnamefont
  {Chou}}, \bibinfo {author} {\bibfnamefont {F.}~\bibnamefont {Wu}}, \bibinfo
  {author} {\bibfnamefont {J.~D.}\ \bibnamefont {Sau}},\ and\ \bibinfo {author}
  {\bibfnamefont {S.}~\bibnamefont {Das~Sarma}},\ }\bibfield  {title} {\bibinfo
  {title} {Acoustic-phonon-mediated superconductivity in rhombohedral trilayer
  graphene},\ }\href {https://doi.org/10.1103/PhysRevLett.127.187001}
  {\bibfield  {journal} {\bibinfo  {journal} {Phys. Rev. Lett.}\ }\textbf
  {\bibinfo {volume} {127}},\ \bibinfo {pages} {187001} (\bibinfo {year}
  {2021})}\BibitemShut {NoStop}%
\bibitem [{\citenamefont {Chou}\ \emph {et~al.}(2022)\citenamefont {Chou},
  \citenamefont {Wu}, \citenamefont {Sau},\ and\ \citenamefont
  {Das~Sarma}}]{Chou_prb22}%
  \BibitemOpen
  \bibfield  {author} {\bibinfo {author} {\bibfnamefont {Y.-Z.}\ \bibnamefont
  {Chou}}, \bibinfo {author} {\bibfnamefont {F.}~\bibnamefont {Wu}}, \bibinfo
  {author} {\bibfnamefont {J.~D.}\ \bibnamefont {Sau}},\ and\ \bibinfo {author}
  {\bibfnamefont {S.}~\bibnamefont {Das~Sarma}},\ }\bibfield  {title} {\bibinfo
  {title} {Acoustic-phonon-mediated superconductivity in bernal bilayer
  graphene},\ }\href {https://doi.org/10.1103/PhysRevB.105.L100503} {\bibfield
  {journal} {\bibinfo  {journal} {Phys. Rev. B}\ }\textbf {\bibinfo {volume}
  {105}},\ \bibinfo {pages} {L100503} (\bibinfo {year} {2022})}\BibitemShut
  {NoStop}%
\bibitem [{\citenamefont {{Firoz Islam}}\ \emph {et~al.}(2022)\citenamefont
  {{Firoz Islam}}, \citenamefont {{Zyuzin}},\ and\ \citenamefont
  {{Zyuzin}}}]{Firoz_cm22}%
  \BibitemOpen
  \bibfield  {author} {\bibinfo {author} {\bibfnamefont {S.}~\bibnamefont
  {{Firoz Islam}}}, \bibinfo {author} {\bibfnamefont {A.~Y.}\ \bibnamefont
  {{Zyuzin}}},\ and\ \bibinfo {author} {\bibfnamefont {A.~A.}\ \bibnamefont
  {{Zyuzin}}},\ }\bibfield  {title} {\bibinfo {title} {{Unconventional
  superconductivity with preformed pairs in twisted bilayer graphene}},\
  }\href@noop {} {\bibfield  {journal} {\bibinfo  {journal} {arXiv e-prints}\
  ,\ \bibinfo {eid} {arXiv:2208.12039}} (\bibinfo {year} {2022})},\ \Eprint
  {https://arxiv.org/abs/2208.12039} {arXiv:2208.12039 [cond-mat.supr-con]}
  \BibitemShut {NoStop}%
\bibitem [{\citenamefont {Lewandowski}\ \emph
  {et~al.}(2021{\natexlab{a}})\citenamefont {Lewandowski}, \citenamefont
  {Chowdhury},\ and\ \citenamefont {Ruhman}}]{Lewandowski_prb21}%
  \BibitemOpen
  \bibfield  {author} {\bibinfo {author} {\bibfnamefont {C.}~\bibnamefont
  {Lewandowski}}, \bibinfo {author} {\bibfnamefont {D.}~\bibnamefont
  {Chowdhury}},\ and\ \bibinfo {author} {\bibfnamefont {J.}~\bibnamefont
  {Ruhman}},\ }\bibfield  {title} {\bibinfo {title} {Pairing in magic-angle
  twisted bilayer graphene: Role of phonon and plasmon umklapp},\ }\href
  {https://doi.org/10.1103/PhysRevB.103.235401} {\bibfield  {journal} {\bibinfo
   {journal} {Phys. Rev. B}\ }\textbf {\bibinfo {volume} {103}},\ \bibinfo
  {pages} {235401} (\bibinfo {year} {2021}{\natexlab{a}})}\BibitemShut
  {NoStop}%
\bibitem [{\citenamefont {Lewandowski}\ \emph
  {et~al.}(2021{\natexlab{b}})\citenamefont {Lewandowski}, \citenamefont
  {Nadj-Perge},\ and\ \citenamefont {Chowdhury}}]{Lewandowski_npj21}%
  \BibitemOpen
  \bibfield  {author} {\bibinfo {author} {\bibfnamefont {C.}~\bibnamefont
  {Lewandowski}}, \bibinfo {author} {\bibfnamefont {S.}~\bibnamefont
  {Nadj-Perge}},\ and\ \bibinfo {author} {\bibfnamefont {D.}~\bibnamefont
  {Chowdhury}},\ }\bibfield  {title} {\bibinfo {title} {Does filling-dependent
  band renormalization aid pairing in twisted bilayer graphene?},\ }\href
  {https://doi.org/10.1038/s41535-021-00379-6} {\bibfield  {journal} {\bibinfo
  {journal} {npj Quantum Materials}\ }\textbf {\bibinfo {volume} {6}},\
  \bibinfo {pages} {82} (\bibinfo {year} {2021}{\natexlab{b}})}\BibitemShut
  {NoStop}%
\bibitem [{\citenamefont {Cea}\ and\ \citenamefont
  {Guinea}(2021)}]{Cea_pnas21}%
  \BibitemOpen
  \bibfield  {author} {\bibinfo {author} {\bibfnamefont {T.}~\bibnamefont
  {Cea}}\ and\ \bibinfo {author} {\bibfnamefont {F.}~\bibnamefont {Guinea}},\
  }\bibfield  {title} {\bibinfo {title} {Coulomb interaction, phonons, and
  superconductivity in twisted bilayer graphene},\ }\href
  {https://doi.org/10.1073/pnas.2107874118} {\bibfield  {journal} {\bibinfo
  {journal} {Proceedings of the National Academy of Sciences}\ }\textbf
  {\bibinfo {volume} {118}},\ \bibinfo {pages} {e2107874118} (\bibinfo {year}
  {2021})},\ \Eprint
  {https://arxiv.org/abs/https://www.pnas.org/doi/pdf/10.1073/pnas.2107874118}
  {https://www.pnas.org/doi/pdf/10.1073/pnas.2107874118} \BibitemShut {NoStop}%
\bibitem [{\citenamefont {Phong}\ \emph {et~al.}(2021)\citenamefont {Phong},
  \citenamefont {Pantale\'on}, \citenamefont {Cea},\ and\ \citenamefont
  {Guinea}}]{phong_prb21}%
  \BibitemOpen
  \bibfield  {author} {\bibinfo {author} {\bibfnamefont {V.~o.~T.}\
  \bibnamefont {Phong}}, \bibinfo {author} {\bibfnamefont {P.~A.}\ \bibnamefont
  {Pantale\'on}}, \bibinfo {author} {\bibfnamefont {T.}~\bibnamefont {Cea}},\
  and\ \bibinfo {author} {\bibfnamefont {F.}~\bibnamefont {Guinea}},\
  }\bibfield  {title} {\bibinfo {title} {Band structure and superconductivity
  in twisted trilayer graphene},\ }\href
  {https://doi.org/10.1103/PhysRevB.104.L121116} {\bibfield  {journal}
  {\bibinfo  {journal} {Phys. Rev. B}\ }\textbf {\bibinfo {volume} {104}},\
  \bibinfo {pages} {L121116} (\bibinfo {year} {2021})}\BibitemShut {NoStop}%
\bibitem [{\citenamefont {Po}\ \emph {et~al.}(2018)\citenamefont {Po},
  \citenamefont {Zou}, \citenamefont {Vishwanath},\ and\ \citenamefont
  {Senthil}}]{Po_prx18}%
  \BibitemOpen
  \bibfield  {author} {\bibinfo {author} {\bibfnamefont {H.~C.}\ \bibnamefont
  {Po}}, \bibinfo {author} {\bibfnamefont {L.}~\bibnamefont {Zou}}, \bibinfo
  {author} {\bibfnamefont {A.}~\bibnamefont {Vishwanath}},\ and\ \bibinfo
  {author} {\bibfnamefont {T.}~\bibnamefont {Senthil}},\ }\bibfield  {title}
  {\bibinfo {title} {Origin of mott insulating behavior and superconductivity
  in twisted bilayer graphene},\ }\href
  {https://doi.org/10.1103/PhysRevX.8.031089} {\bibfield  {journal} {\bibinfo
  {journal} {Phys. Rev. X}\ }\textbf {\bibinfo {volume} {8}},\ \bibinfo {pages}
  {031089} (\bibinfo {year} {2018})}\BibitemShut {NoStop}%
\bibitem [{\citenamefont {{Kozii}}\ \emph {et~al.}(2020)\citenamefont
  {{Kozii}}, \citenamefont {{Zaletel}},\ and\ \citenamefont
  {{Bultinck}}}]{Kozii_cm20}%
  \BibitemOpen
  \bibfield  {author} {\bibinfo {author} {\bibfnamefont {V.}~\bibnamefont
  {{Kozii}}}, \bibinfo {author} {\bibfnamefont {M.~P.}\ \bibnamefont
  {{Zaletel}}},\ and\ \bibinfo {author} {\bibfnamefont {N.}~\bibnamefont
  {{Bultinck}}},\ }\bibfield  {title} {\bibinfo {title} {{Superconductivity in
  a doped valley coherent insulator in magic angle graphene: Goldstone-mediated
  pairing and Kohn-Luttinger mechanism}},\ }\href@noop {} {\bibfield  {journal}
  {\bibinfo  {journal} {arXiv e-prints}\ ,\ \bibinfo {eid} {arXiv:2005.12961}}
  (\bibinfo {year} {2020})},\ \Eprint {https://arxiv.org/abs/2005.12961}
  {arXiv:2005.12961 [cond-mat.str-el]} \BibitemShut {NoStop}%
\bibitem [{\citenamefont {Chatterjee}\ \emph {et~al.}(2022)\citenamefont
  {Chatterjee}, \citenamefont {Wang}, \citenamefont {Berg},\ and\ \citenamefont
  {Zaletel}}]{Chatterjee_natcomm22}%
  \BibitemOpen
  \bibfield  {author} {\bibinfo {author} {\bibfnamefont {S.}~\bibnamefont
  {Chatterjee}}, \bibinfo {author} {\bibfnamefont {T.}~\bibnamefont {Wang}},
  \bibinfo {author} {\bibfnamefont {E.}~\bibnamefont {Berg}},\ and\ \bibinfo
  {author} {\bibfnamefont {M.~P.}\ \bibnamefont {Zaletel}},\ }\bibfield
  {title} {\bibinfo {title} {Inter-valley coherent order and isospin
  fluctuation mediated superconductivity in rhombohedral trilayer graphene},\
  }\href {https://doi.org/10.1038/s41467-022-33561-w} {\bibfield  {journal}
  {\bibinfo  {journal} {Nature Communications}\ }\textbf {\bibinfo {volume}
  {13}},\ \bibinfo {pages} {6013} (\bibinfo {year} {2022})}\BibitemShut
  {NoStop}%
\bibitem [{\citenamefont {{Dong}}\ and\ \citenamefont
  {{Levitov}}(2021{\natexlab{b}})}]{Dong_cm21b}%
  \BibitemOpen
  \bibfield  {author} {\bibinfo {author} {\bibfnamefont {Z.}~\bibnamefont
  {{Dong}}}\ and\ \bibinfo {author} {\bibfnamefont {L.}~\bibnamefont
  {{Levitov}}},\ }\bibfield  {title} {\bibinfo {title} {{Superconductivity in
  the vicinity of an isospin-polarized state in a cubic Dirac band}},\
  }\href@noop {} {\bibfield  {journal} {\bibinfo  {journal} {arXiv e-prints}\
  ,\ \bibinfo {eid} {arXiv:2109.01133}} (\bibinfo {year}
  {2021}{\natexlab{b}})},\ \Eprint {https://arxiv.org/abs/2109.01133}
  {arXiv:2109.01133 [cond-mat.supr-con]} \BibitemShut {NoStop}%
\bibitem [{\citenamefont {{Dong}}\ \emph {et~al.}(2022)\citenamefont {{Dong}},
  \citenamefont {{Chubukov}},\ and\ \citenamefont {{Levitov}}}]{Dong_cm22}%
  \BibitemOpen
  \bibfield  {author} {\bibinfo {author} {\bibfnamefont {Z.}~\bibnamefont
  {{Dong}}}, \bibinfo {author} {\bibfnamefont {A.~V.}\ \bibnamefont
  {{Chubukov}}},\ and\ \bibinfo {author} {\bibfnamefont {L.}~\bibnamefont
  {{Levitov}}},\ }\bibfield  {title} {\bibinfo {title} {{Spin-triplet
  superconductivity at the onset of isospin order in biased bilayer
  graphene}},\ }\href@noop {} {\bibfield  {journal} {\bibinfo  {journal} {arXiv
  e-prints}\ ,\ \bibinfo {eid} {arXiv:2205.13353}} (\bibinfo {year} {2022})},\
  \Eprint {https://arxiv.org/abs/2205.13353} {arXiv:2205.13353
  [cond-mat.supr-con]} \BibitemShut {NoStop}%
\bibitem [{SI()}]{SI}%
  \BibitemOpen
  \href@noop {} {}\bibinfo {note} {See Supplemental Material for: The
  inter-valley scattering from the Coulomb repulsion and the induced Cooper
  pairing; The case of the TBG; The case of the RTG and BBG; Screening of the
  Coulomb interaction.}\BibitemShut {Stop}%
\bibitem [{\citenamefont {Cr{\'e}pel}\ and\ \citenamefont
  {Fu}()}]{crepel_scadv22}%
  \BibitemOpen
  \bibfield  {author} {\bibinfo {author} {\bibfnamefont {V.}~\bibnamefont
  {Cr{\'e}pel}}\ and\ \bibinfo {author} {\bibfnamefont {L.}~\bibnamefont
  {Fu}},\ }\bibfield  {title} {\bibinfo {title} {New mechanism and exact theory
  of superconductivity from strong repulsive interaction},\ }\href
  {https://doi.org/10.1126/sciadv.abh2233} {\bibfield  {journal} {\bibinfo
  {journal} {Science Advances}\ }\textbf {\bibinfo {volume} {7}},\ \bibinfo
  {pages} {eabh2233}}\BibitemShut {NoStop}%
\bibitem [{\citenamefont {Crépel}\ and\ \citenamefont
  {Fu}(2022)}]{crepel_pnas22}%
  \BibitemOpen
  \bibfield  {author} {\bibinfo {author} {\bibfnamefont {V.}~\bibnamefont
  {Crépel}}\ and\ \bibinfo {author} {\bibfnamefont {L.}~\bibnamefont {Fu}},\
  }\bibfield  {title} {\bibinfo {title} {Spin-triplet superconductivity from
  excitonic effect in doped insulators},\ }\href
  {https://doi.org/10.1073/pnas.2117735119} {\bibfield  {journal} {\bibinfo
  {journal} {Proceedings of the National Academy of Sciences}\ }\textbf
  {\bibinfo {volume} {119}},\ \bibinfo {pages} {e2117735119} (\bibinfo {year}
  {2022})},\ \Eprint
  {https://arxiv.org/abs/https://www.pnas.org/doi/pdf/10.1073/pnas.2117735119}
  {https://www.pnas.org/doi/pdf/10.1073/pnas.2117735119} \BibitemShut {NoStop}%
\bibitem [{\citenamefont {Guinea}\ and\ \citenamefont
  {Uchoa}(2012)}]{GuineaUchoa_prb12}%
  \BibitemOpen
  \bibfield  {author} {\bibinfo {author} {\bibfnamefont {F.}~\bibnamefont
  {Guinea}}\ and\ \bibinfo {author} {\bibfnamefont {B.}~\bibnamefont {Uchoa}},\
  }\bibfield  {title} {\bibinfo {title} {Odd-momentum pairing and
  superconductivity in vertical graphene heterostructures},\ }\href
  {https://doi.org/10.1103/PhysRevB.86.134521} {\bibfield  {journal} {\bibinfo
  {journal} {Phys. Rev. B}\ }\textbf {\bibinfo {volume} {86}},\ \bibinfo
  {pages} {134521} (\bibinfo {year} {2012})}\BibitemShut {NoStop}%
\bibitem [{\citenamefont {Rold\'an}\ \emph {et~al.}(2013)\citenamefont
  {Rold\'an}, \citenamefont {Cappelluti},\ and\ \citenamefont
  {Guinea}}]{Roldan_prb13}%
  \BibitemOpen
  \bibfield  {author} {\bibinfo {author} {\bibfnamefont {R.}~\bibnamefont
  {Rold\'an}}, \bibinfo {author} {\bibfnamefont {E.}~\bibnamefont
  {Cappelluti}},\ and\ \bibinfo {author} {\bibfnamefont {F.}~\bibnamefont
  {Guinea}},\ }\bibfield  {title} {\bibinfo {title} {Interactions and
  superconductivity in heavily doped mos${}_{2}$},\ }\href
  {https://doi.org/10.1103/PhysRevB.88.054515} {\bibfield  {journal} {\bibinfo
  {journal} {Phys. Rev. B}\ }\textbf {\bibinfo {volume} {88}},\ \bibinfo
  {pages} {054515} (\bibinfo {year} {2013})}\BibitemShut {NoStop}%
\bibitem [{\citenamefont {Parr}\ \emph {et~al.}(1950)\citenamefont {Parr},
  \citenamefont {Craig},\ and\ \citenamefont {Ross}}]{Parr_JCP1950}%
  \BibitemOpen
  \bibfield  {author} {\bibinfo {author} {\bibfnamefont {R.~G.}\ \bibnamefont
  {Parr}}, \bibinfo {author} {\bibfnamefont {D.~P.}\ \bibnamefont {Craig}},\
  and\ \bibinfo {author} {\bibfnamefont {I.~G.}\ \bibnamefont {Ross}},\
  }\bibfield  {title} {\bibinfo {title} {Molecular orbital calculations of the
  lower excited electronic levels of benzene, configuration interaction
  included},\ }\href {https://doi.org/10.1063/1.1747540} {\bibfield  {journal}
  {\bibinfo  {journal} {The Journal of Chemical Physics}\ }\textbf {\bibinfo
  {volume} {18}},\ \bibinfo {pages} {1561} (\bibinfo {year} {1950})},\ \Eprint
  {https://arxiv.org/abs/https://doi.org/10.1063/1.1747540}
  {https://doi.org/10.1063/1.1747540} \BibitemShut {NoStop}%
\bibitem [{\citenamefont {Ohno}(1964)}]{Ohno_TCA1964}%
  \BibitemOpen
  \bibfield  {author} {\bibinfo {author} {\bibfnamefont {K.}~\bibnamefont
  {Ohno}},\ }\bibfield  {title} {\bibinfo {title} {Some remarks on the
  pariser-parr-pople method},\ }\href {https://doi.org/10.1007/BF00528281}
  {\bibfield  {journal} {\bibinfo  {journal} {Theoretica chimica acta}\
  }\textbf {\bibinfo {volume} {2}},\ \bibinfo {pages} {219} (\bibinfo {year}
  {1964})}\BibitemShut {NoStop}%
\bibitem [{\citenamefont {Baird}\ and\ \citenamefont
  {Dewar}(1969)}]{Baird_JCP1969}%
  \BibitemOpen
  \bibfield  {author} {\bibinfo {author} {\bibfnamefont {N.~C.}\ \bibnamefont
  {Baird}}\ and\ \bibinfo {author} {\bibfnamefont {M.~J.~S.}\ \bibnamefont
  {Dewar}},\ }\bibfield  {title} {\bibinfo {title} {Ground states of
  $\sigma$-bonded molecules. iv. the mindo method and its application to
  hydrocarbons},\ }\href {https://doi.org/10.1063/1.1671186} {\bibfield
  {journal} {\bibinfo  {journal} {The Journal of Chemical Physics}\ }\textbf
  {\bibinfo {volume} {50}},\ \bibinfo {pages} {1262} (\bibinfo {year}
  {1969})},\ \Eprint {https://arxiv.org/abs/https://doi.org/10.1063/1.1671186}
  {https://doi.org/10.1063/1.1671186} \BibitemShut {NoStop}%
\bibitem [{\citenamefont {Verg\'es}\ \emph {et~al.}(2010)\citenamefont
  {Verg\'es}, \citenamefont {SanFabi\'an}, \citenamefont {Chiappe},\ and\
  \citenamefont {Louis}}]{Verges_prb10}%
  \BibitemOpen
  \bibfield  {author} {\bibinfo {author} {\bibfnamefont {J.~A.}\ \bibnamefont
  {Verg\'es}}, \bibinfo {author} {\bibfnamefont {E.}~\bibnamefont
  {SanFabi\'an}}, \bibinfo {author} {\bibfnamefont {G.}~\bibnamefont
  {Chiappe}},\ and\ \bibinfo {author} {\bibfnamefont {E.}~\bibnamefont
  {Louis}},\ }\bibfield  {title} {\bibinfo {title} {Fit of pariser-parr-pople
  and hubbard model hamiltonians to charge and spin states of polycyclic
  aromatic hydrocarbons},\ }\href {https://doi.org/10.1103/PhysRevB.81.085120}
  {\bibfield  {journal} {\bibinfo  {journal} {Phys. Rev. B}\ }\textbf {\bibinfo
  {volume} {81}},\ \bibinfo {pages} {085120} (\bibinfo {year}
  {2010})}\BibitemShut {NoStop}%
\bibitem [{\citenamefont {Wehling}\ \emph {et~al.}(2011)\citenamefont
  {Wehling}, \citenamefont {\ifmmode \mbox{\c{S}}\else \c{S}\fi{}a\ifmmode
  \mbox{\c{s}}\else \c{s}\fi{}\ifmmode \imath \else \i
  \fi{}o\ifmmode~\breve{g}\else \u{g}\fi{}lu}, \citenamefont {Friedrich},
  \citenamefont {Lichtenstein}, \citenamefont {Katsnelson},\ and\ \citenamefont
  {Bl\"ugel}}]{Wehling_prl11}%
  \BibitemOpen
  \bibfield  {author} {\bibinfo {author} {\bibfnamefont {T.~O.}\ \bibnamefont
  {Wehling}}, \bibinfo {author} {\bibfnamefont {E.}~\bibnamefont {\ifmmode
  \mbox{\c{S}}\else \c{S}\fi{}a\ifmmode \mbox{\c{s}}\else \c{s}\fi{}\ifmmode
  \imath \else \i \fi{}o\ifmmode~\breve{g}\else \u{g}\fi{}lu}}, \bibinfo
  {author} {\bibfnamefont {C.}~\bibnamefont {Friedrich}}, \bibinfo {author}
  {\bibfnamefont {A.~I.}\ \bibnamefont {Lichtenstein}}, \bibinfo {author}
  {\bibfnamefont {M.~I.}\ \bibnamefont {Katsnelson}},\ and\ \bibinfo {author}
  {\bibfnamefont {S.}~\bibnamefont {Bl\"ugel}},\ }\bibfield  {title} {\bibinfo
  {title} {Strength of effective coulomb interactions in graphene and
  graphite},\ }\href {https://doi.org/10.1103/PhysRevLett.106.236805}
  {\bibfield  {journal} {\bibinfo  {journal} {Phys. Rev. Lett.}\ }\textbf
  {\bibinfo {volume} {106}},\ \bibinfo {pages} {236805} (\bibinfo {year}
  {2011})}\BibitemShut {NoStop}%
\bibitem [{\citenamefont {Lopes~dos Santos}\ \emph {et~al.}(2007)\citenamefont
  {Lopes~dos Santos}, \citenamefont {Peres},\ and\ \citenamefont
  {Castro~Neto}}]{dossantos_prl07}%
  \BibitemOpen
  \bibfield  {author} {\bibinfo {author} {\bibfnamefont {J.~M.~B.}\
  \bibnamefont {Lopes~dos Santos}}, \bibinfo {author} {\bibfnamefont
  {N.~M.~R.}\ \bibnamefont {Peres}},\ and\ \bibinfo {author} {\bibfnamefont
  {A.~H.}\ \bibnamefont {Castro~Neto}},\ }\bibfield  {title} {\bibinfo {title}
  {Graphene bilayer with a twist: Electronic structure},\ }\href
  {https://doi.org/10.1103/PhysRevLett.99.256802} {\bibfield  {journal}
  {\bibinfo  {journal} {Phys. Rev. Lett.}\ }\textbf {\bibinfo {volume} {99}},\
  \bibinfo {pages} {256802} (\bibinfo {year} {2007})}\BibitemShut {NoStop}%
\bibitem [{\citenamefont {Bistritzer}\ and\ \citenamefont
  {MacDonald}(2011)}]{bistritzer_pnas11}%
  \BibitemOpen
  \bibfield  {author} {\bibinfo {author} {\bibfnamefont {R.}~\bibnamefont
  {Bistritzer}}\ and\ \bibinfo {author} {\bibfnamefont {A.~H.}\ \bibnamefont
  {MacDonald}},\ }\bibfield  {title} {\bibinfo {title} {Moir\&\#xe9; bands in
  twisted double-layer graphene},\ }\href
  {https://doi.org/10.1073/pnas.1108174108} {\bibfield  {journal} {\bibinfo
  {journal} {Proceedings of the National Academy of Sciences}\ }\textbf
  {\bibinfo {volume} {108}},\ \bibinfo {pages} {12233} (\bibinfo {year}
  {2011})},\ \Eprint
  {https://arxiv.org/abs/https://www.pnas.org/doi/pdf/10.1073/pnas.1108174108}
  {https://www.pnas.org/doi/pdf/10.1073/pnas.1108174108} \BibitemShut {NoStop}%
\bibitem [{\citenamefont {Lopes~dos Santos}\ \emph {et~al.}(2012)\citenamefont
  {Lopes~dos Santos}, \citenamefont {Peres},\ and\ \citenamefont
  {Castro~Neto}}]{dossantos_prb12}%
  \BibitemOpen
  \bibfield  {author} {\bibinfo {author} {\bibfnamefont {J.~M.~B.}\
  \bibnamefont {Lopes~dos Santos}}, \bibinfo {author} {\bibfnamefont
  {N.~M.~R.}\ \bibnamefont {Peres}},\ and\ \bibinfo {author} {\bibfnamefont
  {A.~H.}\ \bibnamefont {Castro~Neto}},\ }\bibfield  {title} {\bibinfo {title}
  {Continuum model of the twisted graphene bilayer},\ }\href
  {https://doi.org/10.1103/PhysRevB.86.155449} {\bibfield  {journal} {\bibinfo
  {journal} {Phys. Rev. B}\ }\textbf {\bibinfo {volume} {86}},\ \bibinfo
  {pages} {155449} (\bibinfo {year} {2012})}\BibitemShut {NoStop}%
\bibitem [{\citenamefont {Koshino}\ \emph {et~al.}(2018)\citenamefont
  {Koshino}, \citenamefont {Yuan}, \citenamefont {Koretsune}, \citenamefont
  {Ochi}, \citenamefont {Kuroki},\ and\ \citenamefont {Fu}}]{koshino_prx18}%
  \BibitemOpen
  \bibfield  {author} {\bibinfo {author} {\bibfnamefont {M.}~\bibnamefont
  {Koshino}}, \bibinfo {author} {\bibfnamefont {N.~F.~Q.}\ \bibnamefont
  {Yuan}}, \bibinfo {author} {\bibfnamefont {T.}~\bibnamefont {Koretsune}},
  \bibinfo {author} {\bibfnamefont {M.}~\bibnamefont {Ochi}}, \bibinfo {author}
  {\bibfnamefont {K.}~\bibnamefont {Kuroki}},\ and\ \bibinfo {author}
  {\bibfnamefont {L.}~\bibnamefont {Fu}},\ }\bibfield  {title} {\bibinfo
  {title} {Maximally localized wannier orbitals and the extended hubbard model
  for twisted bilayer graphene},\ }\href
  {https://doi.org/10.1103/PhysRevX.8.031087} {\bibfield  {journal} {\bibinfo
  {journal} {Phys. Rev. X}\ }\textbf {\bibinfo {volume} {8}},\ \bibinfo {pages}
  {031087} (\bibinfo {year} {2018})}\BibitemShut {NoStop}%
\bibitem [{\citenamefont {Zibrov}\ \emph {et~al.}(2018)\citenamefont {Zibrov},
  \citenamefont {Rao}, \citenamefont {Kometter}, \citenamefont {Spanton},
  \citenamefont {Li}, \citenamefont {Dean}, \citenamefont {Taniguchi},
  \citenamefont {Watanabe}, \citenamefont {Serbyn},\ and\ \citenamefont
  {Young}}]{zibrov_prl18}%
  \BibitemOpen
  \bibfield  {author} {\bibinfo {author} {\bibfnamefont {A.~A.}\ \bibnamefont
  {Zibrov}}, \bibinfo {author} {\bibfnamefont {P.}~\bibnamefont {Rao}},
  \bibinfo {author} {\bibfnamefont {C.}~\bibnamefont {Kometter}}, \bibinfo
  {author} {\bibfnamefont {E.~M.}\ \bibnamefont {Spanton}}, \bibinfo {author}
  {\bibfnamefont {J.~I.~A.}\ \bibnamefont {Li}}, \bibinfo {author}
  {\bibfnamefont {C.~R.}\ \bibnamefont {Dean}}, \bibinfo {author}
  {\bibfnamefont {T.}~\bibnamefont {Taniguchi}}, \bibinfo {author}
  {\bibfnamefont {K.}~\bibnamefont {Watanabe}}, \bibinfo {author}
  {\bibfnamefont {M.}~\bibnamefont {Serbyn}},\ and\ \bibinfo {author}
  {\bibfnamefont {A.~F.}\ \bibnamefont {Young}},\ }\bibfield  {title} {\bibinfo
  {title} {Emergent dirac gullies and gully-symmetry-breaking quantum hall
  states in $aba$ trilayer graphene},\ }\href
  {https://doi.org/10.1103/PhysRevLett.121.167601} {\bibfield  {journal}
  {\bibinfo  {journal} {Phys. Rev. Lett.}\ }\textbf {\bibinfo {volume} {121}},\
  \bibinfo {pages} {167601} (\bibinfo {year} {2018})}\BibitemShut {NoStop}%
\bibitem [{\citenamefont {Zhou}\ \emph
  {et~al.}(2021{\natexlab{b}})\citenamefont {Zhou}, \citenamefont {Xie},
  \citenamefont {Ghazaryan}, \citenamefont {Holder}, \citenamefont {Ehrets},
  \citenamefont {Spanton}, \citenamefont {Taniguchi}, \citenamefont {Watanabe},
  \citenamefont {Berg}, \citenamefont {Serbyn},\ and\ \citenamefont
  {Young}}]{Zhou_nature21_bis}%
  \BibitemOpen
  \bibfield  {author} {\bibinfo {author} {\bibfnamefont {H.}~\bibnamefont
  {Zhou}}, \bibinfo {author} {\bibfnamefont {T.}~\bibnamefont {Xie}}, \bibinfo
  {author} {\bibfnamefont {A.}~\bibnamefont {Ghazaryan}}, \bibinfo {author}
  {\bibfnamefont {T.}~\bibnamefont {Holder}}, \bibinfo {author} {\bibfnamefont
  {J.~R.}\ \bibnamefont {Ehrets}}, \bibinfo {author} {\bibfnamefont {E.~M.}\
  \bibnamefont {Spanton}}, \bibinfo {author} {\bibfnamefont {T.}~\bibnamefont
  {Taniguchi}}, \bibinfo {author} {\bibfnamefont {K.}~\bibnamefont {Watanabe}},
  \bibinfo {author} {\bibfnamefont {E.}~\bibnamefont {Berg}}, \bibinfo {author}
  {\bibfnamefont {M.}~\bibnamefont {Serbyn}},\ and\ \bibinfo {author}
  {\bibfnamefont {A.~F.}\ \bibnamefont {Young}},\ }\bibfield  {title} {\bibinfo
  {title} {Half- and quarter-metals in rhombohedral trilayer graphene},\ }\href
  {https://doi.org/10.1038/s41586-021-03938-w} {\bibfield  {journal} {\bibinfo
  {journal} {Nature}\ }\textbf {\bibinfo {volume} {598}},\ \bibinfo {pages}
  {429} (\bibinfo {year} {2021}{\natexlab{b}})}\BibitemShut {NoStop}%
\bibitem [{\citenamefont {Zhang}\ \emph {et~al.}(2010)\citenamefont {Zhang},
  \citenamefont {Sahu}, \citenamefont {Min},\ and\ \citenamefont
  {MacDonald}}]{Zhang_prb10}%
  \BibitemOpen
  \bibfield  {author} {\bibinfo {author} {\bibfnamefont {F.}~\bibnamefont
  {Zhang}}, \bibinfo {author} {\bibfnamefont {B.}~\bibnamefont {Sahu}},
  \bibinfo {author} {\bibfnamefont {H.}~\bibnamefont {Min}},\ and\ \bibinfo
  {author} {\bibfnamefont {A.~H.}\ \bibnamefont {MacDonald}},\ }\bibfield
  {title} {\bibinfo {title} {Band structure of $abc$-stacked graphene
  trilayers},\ }\href {https://doi.org/10.1103/PhysRevB.82.035409} {\bibfield
  {journal} {\bibinfo  {journal} {Phys. Rev. B}\ }\textbf {\bibinfo {volume}
  {82}},\ \bibinfo {pages} {035409} (\bibinfo {year} {2010})}\BibitemShut
  {NoStop}%
\bibitem [{\citenamefont {Nakamura}\ \emph {et~al.}(2018)\citenamefont
  {Nakamura}, \citenamefont {Kim}, \citenamefont {Ichinokura}, \citenamefont
  {Takayama}, \citenamefont {Zotov}, \citenamefont {Saranin}, \citenamefont
  {Hasegawa},\ and\ \citenamefont {Hasegawa}}]{Nakamura_prb18}%
  \BibitemOpen
  \bibfield  {author} {\bibinfo {author} {\bibfnamefont {T.}~\bibnamefont
  {Nakamura}}, \bibinfo {author} {\bibfnamefont {H.}~\bibnamefont {Kim}},
  \bibinfo {author} {\bibfnamefont {S.}~\bibnamefont {Ichinokura}}, \bibinfo
  {author} {\bibfnamefont {A.}~\bibnamefont {Takayama}}, \bibinfo {author}
  {\bibfnamefont {A.~V.}\ \bibnamefont {Zotov}}, \bibinfo {author}
  {\bibfnamefont {A.~A.}\ \bibnamefont {Saranin}}, \bibinfo {author}
  {\bibfnamefont {Y.}~\bibnamefont {Hasegawa}},\ and\ \bibinfo {author}
  {\bibfnamefont {S.}~\bibnamefont {Hasegawa}},\ }\bibfield  {title} {\bibinfo
  {title} {Unconventional superconductivity in the single-atom-layer alloy
  $\mathrm{Si}$(111)-$\sqrt{3}\times\sqrt{3}$-($\mathrm{Tl,Pb}$)},\ }\href
  {https://doi.org/10.1103/PhysRevB.98.134505} {\bibfield  {journal} {\bibinfo
  {journal} {Phys. Rev. B}\ }\textbf {\bibinfo {volume} {98}},\ \bibinfo
  {pages} {134505} (\bibinfo {year} {2018})}\BibitemShut {NoStop}%
\bibitem [{\citenamefont {Wu}\ \emph {et~al.}(2020)\citenamefont {Wu},
  \citenamefont {Ming}, \citenamefont {Smith}, \citenamefont {Liu},
  \citenamefont {Ye}, \citenamefont {Wang}, \citenamefont {Johnston},\ and\
  \citenamefont {Weitering}}]{Wu_prl20}%
  \BibitemOpen
  \bibfield  {author} {\bibinfo {author} {\bibfnamefont {X.}~\bibnamefont
  {Wu}}, \bibinfo {author} {\bibfnamefont {F.}~\bibnamefont {Ming}}, \bibinfo
  {author} {\bibfnamefont {T.~S.}\ \bibnamefont {Smith}}, \bibinfo {author}
  {\bibfnamefont {G.}~\bibnamefont {Liu}}, \bibinfo {author} {\bibfnamefont
  {F.}~\bibnamefont {Ye}}, \bibinfo {author} {\bibfnamefont {K.}~\bibnamefont
  {Wang}}, \bibinfo {author} {\bibfnamefont {S.}~\bibnamefont {Johnston}},\
  and\ \bibinfo {author} {\bibfnamefont {H.~H.}\ \bibnamefont {Weitering}},\
  }\bibfield  {title} {\bibinfo {title} {Superconductivity in a hole-doped
  mott-insulating triangular adatom layer on a silicon surface},\ }\href
  {https://doi.org/10.1103/PhysRevLett.125.117001} {\bibfield  {journal}
  {\bibinfo  {journal} {Phys. Rev. Lett.}\ }\textbf {\bibinfo {volume} {125}},\
  \bibinfo {pages} {117001} (\bibinfo {year} {2020})}\BibitemShut {NoStop}%
\bibitem [{\citenamefont {{Ming}}\ \emph {et~al.}(2022)\citenamefont {{Ming}},
  \citenamefont {{Wu}}, \citenamefont {{Chen}}, \citenamefont {{Wang}},
  \citenamefont {{Mai}}, \citenamefont {{Maier}}, \citenamefont {{Strockoz}},
  \citenamefont {{Venderbos}}, \citenamefont {{Gonzalez}}, \citenamefont
  {{Ortega}}, \citenamefont {{Johnston}},\ and\ \citenamefont
  {{Weitering}}}]{Ming_cm22}%
  \BibitemOpen
  \bibfield  {author} {\bibinfo {author} {\bibfnamefont {F.}~\bibnamefont
  {{Ming}}}, \bibinfo {author} {\bibfnamefont {X.}~\bibnamefont {{Wu}}},
  \bibinfo {author} {\bibfnamefont {C.}~\bibnamefont {{Chen}}}, \bibinfo
  {author} {\bibfnamefont {K.~D.}\ \bibnamefont {{Wang}}}, \bibinfo {author}
  {\bibfnamefont {P.}~\bibnamefont {{Mai}}}, \bibinfo {author} {\bibfnamefont
  {T.~A.}\ \bibnamefont {{Maier}}}, \bibinfo {author} {\bibfnamefont
  {J.}~\bibnamefont {{Strockoz}}}, \bibinfo {author} {\bibfnamefont {J.~W.~F.}\
  \bibnamefont {{Venderbos}}}, \bibinfo {author} {\bibfnamefont
  {C.}~\bibnamefont {{Gonzalez}}}, \bibinfo {author} {\bibfnamefont
  {J.}~\bibnamefont {{Ortega}}}, \bibinfo {author} {\bibfnamefont
  {S.}~\bibnamefont {{Johnston}}},\ and\ \bibinfo {author} {\bibfnamefont
  {H.~H.}\ \bibnamefont {{Weitering}}},\ }\bibfield  {title} {\bibinfo {title}
  {{Evidence for chiral superconductivity on a silicon surface}},\ }\href@noop
  {} {\bibfield  {journal} {\bibinfo  {journal} {arXiv e-prints}\ ,\ \bibinfo
  {eid} {arXiv:2210.06273}} (\bibinfo {year} {2022})},\ \Eprint
  {https://arxiv.org/abs/2210.06273} {arXiv:2210.06273 [cond-mat.supr-con]}
  \BibitemShut {NoStop}%
\end{thebibliography}%


\clearpage

\onecolumngrid

\setcounter{section}{0}
\setcounter{equation}{0}
\setcounter{figure}{0}
\setcounter{table}{0}
\setcounter{page}{1}
\makeatletter
\renewcommand{\theequation}{S\arabic{equation}}
\renewcommand{\thefigure}{S\arabic{figure}}

\begin{center}
\Large Supplementary information for:\\
\textbf{Superconductivity induced by the inter-valley Coulomb scattering in few layers of graphene}
\end{center}


\section{The inter-valley scattering from the Coulomb repulsion
and the induced Cooper pairing}

We consider the Hamiltonian of the Coulomb repulsion between the $p_z$
electrons within the minimal lattice model for a multilayer of graphene:
\bea\label{SI:H_coulomb}
\hat{H}_{int}=\frac{1}{2}
\sum_{\vec{R}\vec{R}'}
\sum_{ij\sigma\sigma'}
c^{\dagger}_{i\sigma}(\vec{R})c^{\dagger}_{j\sigma'}(\vec{R}')
V^{ij}_{C}(\vec{R}-\vec{R}')
c_{j\sigma'}(\vec{R}')c_{i\sigma}(\vec{R}),
\eea
where $\vec{R}$ are the coordinates of the Bravais lattice, $i,j$ are the labels of the sub-lattice/layer,
$c_{i\sigma}(\vec{R})$ is the the quantum operator for the annihilation
of one electron with spin $\sigma$ in the $p_z$ orbital localized at the position $\vec{R}+\vec{\delta}_i$,
with $\vec{\delta}_i$ the internal coordinate in the unit cell,
and:
\bea
V^{ij}_C(\vec{R}-\vec{R}')=\frac{e^2}{4\pi\epsilon\left| 
\vec{R}-\vec{R}'+\vec{\delta}_i-\vec{\delta}_j
\right|}
\eea
is the Coulomb potential, where $e$ is the electron charge,
$\epsilon$ is the dielectric constant of the environment,
$\epsilon=\epsilon_0$ in the vacuum.
Next, we consider the continuum limit of the lattice model,
by expanding the operators $c$ as:
\bea\label{SI:continuum_fields}
c_{i\sigma}(\vec{R})\equiv A_c^{1/2}\left[
\psi^{K}_{i\sigma}(\vec{R})e^{iK\cdot\vec{R}}+
\psi^{K'}_{i\sigma}(\vec{R})e^{iK'\cdot\vec{R}}
\right],
\eea
where $A_c=\sqrt{3}a^2/2$ is the area of the unit cell of graphene,
$a=2.46${\AA} being the lattice constant,
 $K,K'$ are the non equivalent corners of the BZ
and $\psi^{K}_{i\sigma}(\vec{r}),\psi^{K'}_{i\sigma}(\vec{r})$
are fermionic operators, which vary smoothly with the continuum position, $\vec{r}$,
and represent the valley projections of $c_{i\sigma}(\vec{R})$.
Replacing the Eq. \pref{SI:continuum_fields} into the Eq. \pref{SI:H_coulomb},
among all the terms one finds the following valley-exchange interaction:
\bea\label{SI:Hexc}
\hat{H}_{exc}=
A_c^2
\sum_{\vec{R}\vec{R}'}
\sum_{ij\sigma\sigma'}
\psi^{K,\dagger}_{i\sigma}(\vec{R})\psi^{K',\dagger}_{j\sigma'}(\vec{R}')
\psi^{K}_{j\sigma'}(\vec{R}')\psi^{K'}_{i\sigma}(\vec{R})
V^{ij}_{C}(\vec{R}-\vec{R}')e^{-i\Delta K\cdot(\vec{R}-\vec{R}')},
\eea
which describes the inter-valley scattering processes.
Note that we used:
$V^{ij}_{C}(\vec{R}-\vec{R}')=V^{ji}_{C}(\vec{R}'-\vec{R})$,
in order to remove the factor $1/2$ ahead of the Eq. \pref{SI:Hexc}.
We exploit the fact that the Fourier envelope of $V_{C}$ varies slowly for wave vectors
close to $\Delta K$
to assume that the interaction potential of the Eq. \pref{SI:Hexc} is approximately local:
\bea\label{SI:local}
V^{ij}_{C}(\vec{R}-\vec{R}')e^{-i\Delta K\cdot(\vec{R}-\vec{R}')}\simeq
A_c^{-1}v^{ij}_{C}\left(\Delta K\right)\delta_{\vec{R}\vec{R}'},
\eea
where:
\bea\label{SI:v_C_def}
v^{ij}_{C}\left(\Delta K\right)\equiv A_c\sum_{\vec{R}}
V^{ij}_{C}(\vec{R}-\vec{R}')e^{-i\Delta K\cdot(\vec{R}-\vec{R}')}.
\eea 
Using: $A_c\sum_{\vec{R}}\simeq \int\,d^2\vec{r}$ within the previous expression,
one obtains:
\bea\label{SI:J_def}
v^{ij}_{C}\left(\Delta K\right)\simeq
\delta^{ij}J\quad,\quad J\equiv\frac{e^2}{2\epsilon|\Delta K|},
\eea
where the elements in which $i,j$ correspond to the same layer
but to different sub-lattices vanish exactly by symmetry (see the App. \ref{SI:app:V_DK_symm}),
while we neglected the elements in which $i,j$ correspond to different layers,
that are suppressed by the factor
$e^{-\left|\Delta K\right| d_{ij}}$,
$d_{ij}>0$ being the interlayer distance.
Using the Eqs. \pref{SI:local} and \pref{SI:J_def} into the Eq. \pref{SI:Hexc} and
taking the continuum limit finally gives:
\bea
\hat{H}_{exc}&\simeq&
J
\sum_{i\sigma\sigma'}
\int\,d^2\vec{r}
\psi^{K,\dagger}_{i\sigma}(\vec{r})\psi^{K',\dagger}_{i\sigma'}(\vec{r})
\psi^{K}_{i\sigma'}(\vec{r})\psi^{K'}_{i\sigma}(\vec{r})=\nn\\
&=&-
J\sum_{i\sigma\sigma'}
\int\,d^2\vec{r}
\psi^{K,\dagger}_{i\sigma}(\vec{r})\psi^{K',\dagger}_{i\sigma'}(\vec{r})
\psi^{K'}_{i\sigma}(\vec{r})\psi^{K}_{i\sigma'}(\vec{r}),\label{SI:Hexc_cont}
\eea
where we have permuted two operators in the second line,
picking up the minus sign.
Because $J>0$, the Eq. \pref{SI:Hexc_cont} provides an attractive interaction
favoring the local occupancy of the two valleys and hence
the Cooper pairing with electrons in opposite valleys.
In order to figure out the spin
structure
of the Cooper pairs
allowed by the interaction of the Eq. \pref{SI:Hexc_cont},
it is convenient to define the following two-fermions operators:
\begin{subequations}
\bea\label{SI:triplet}
\Psi_{1,-1}^{i}(\vec{r})=\psi^{K'}_{i\downarrow}(\vec{r})\psi^{K}_{i\downarrow}(\vec{r})\quad,\quad
\Psi_{1,0}^{i}(\vec{r})&=&\frac{\psi^{K'}_{i\downarrow}(\vec{r})\psi^{K}_{i\uparrow}(\vec{r})+
\psi^{K'}_{i\uparrow}(\vec{r})\psi^{K}_{i\downarrow}(\vec{r})}{\sqrt{2}}\quad,\quad
\Psi_{1,1}^{i}(\vec{r})=\psi^{K'}_{i\uparrow}(\vec{r})\psi^{K}_{i\uparrow}(\vec{r}),
\eea
each of those annihilates a spin triplet Cooper pair, and:
\bea\label{SI:singlet}
\Psi_{0,0}^{i}(\vec{r})&=&\frac{\psi^{K'}_{i\downarrow}(\vec{r})\psi^{K}_{i\uparrow}(\vec{r})-
\psi^{K'}_{i\uparrow}(\vec{r})\psi^{K}_{i\downarrow}(\vec{r})}{\sqrt{2}},
\eea
which in turns annihilates a spin singlet Cooper pair.
\end{subequations}
Then the Eq. \pref{SI:Hexc_cont} can be written as:
\bea
\hat{H}_{exc}\simeq-J\sum_{i}\int\,d^2\vec{r}
\left[
\sum_{s=-1,0,1}\Psi_{1,s}^{i,\dagger}(\vec{r})\Psi_{1,s}^{i}(\vec{r})-
\Psi_{0,0}^{i,\dagger}(\vec{r})\Psi_{0,0}^{i}(\vec{r})
\right],
\eea
which is attractive in the triplet channel and repulsive in the singlet one,
meaning that only the spin triplet Cooper pairs are stable and,
in addition, they are degenerate.
The SC OP is purely local:
\bea
\Delta^i(\vec{r})\propto\left\langle
\Psi_{1,s}^{i}(\vec{r})
  \right\rangle.
\eea
Within the BCS theory,
the value of $T_c$ can be obtained by looking
for nonzero solutions of the linearized gap equation:
\bea\label{SI:linearized_gap_eq}
\Delta^i(\vec{r})=\frac{J}{\beta}\sum_j\int\,d\vec{r}'
\sum_{l=-\infty}^{+\infty}
\mathcal{G}^K_{ij}\left(\vec{r},\vec{r}';i\omega_l\right)
\mathcal{G}^{K'}_{ij}\left(\vec{r},\vec{r}';-i\omega_l\right)
\Delta^j\left(\vec{r}'\right),
\eea 
where $\beta=(K_BT)^{-1}$ is the inverse of the temperature,
$\omega_l=\pi (2l+1)/(\hbar\beta)$ are fermionic Matsubara frequencies
and $\mathcal{G}^{K,K'}$ are the Green's function for the $K,K'$ valleys,
respectively, computed in the normal phase.
The Eq. \pref{SI:linearized_gap_eq} is written in real space in order to be as general as possible,
holding also for the systems that are not translationally invariant,
as is the case in the presence of a moir\'e.
In all the cases preserving the translational invariance, then the OP is uniform:
$\Delta^i(\vec{r})\equiv \Delta^i$, and the Eq. \pref{SI:linearized_gap_eq}
reduces to the familiar expression:
\bea\label{SI:linearized_gap_eq_TI}
\Delta^i=\frac{J}{\beta\Omega}
\sum_j
\sum_{\vec{q}}
\sum_{l=-\infty}^{+\infty}
\mathcal{G}^K_{ij}\left(\vec{q},i\omega_l\right)
\mathcal{G}^{K'}_{ij}\left(-\vec{q},-i\omega_l\right)
\Delta^j,
\eea 
where $\Omega$ is the area of the system.

\section{The case of the TBG}
We consider the continuum model of the TBG
introduced in the Refs. \cite{dossantos_prl07,bistritzer_pnas11},
that describes the hopping between the electrons in opposite layers
by means of a local potential having the periodicity of the moir\'e.
This potential breaks the translational invariance within each moir\'e unit cell and strongly
hybridizes the $p_z$
orbitals of the constitutive graphene sheets,
generating a number of weakly dispersive minibands at small twist angles.
The Bloch's eigenfunction corresponding to the band $E^K_{n,\vec{k}}$ of the $K$-valley
can be written as:
\bea
\Phi^K_{i,n,\vec{k}}(\vec{r})=\frac{e^{i\vec{k}\cdot\vec{r}}}{\sqrt{\Omega}}
\sum_{\vec{G}}e^{i\vec{G}\cdot\vec{r}}\phi^K_{i,n,\vec{k}}(\vec{G}),
\eea
where $\vec{k}$ is the wave-vector in the moir\'e BZ,
the $\vec{G}$'s are the vectors of the moir\'e reciprocal lattice
and the $\phi^K_{n,\vec{k}}(\vec{G})$'s are numeric eigenvectors normalized according to:
$\sum_{\vec{G}i}\phi^{K,*}_{i,n,\vec{k}}(\vec{G})\phi^K_{i,m,\vec{k}}(\vec{G})=\delta_{nm}$.
The energies and eigenfunctions corresponding to the $K'$
valley are related to the ones at $K$ by the $\mathcal{C}_2\mathcal{T}$ symmetry:
\bea\label{SI:C2T_def}
E^{K'}_{n,\vec{k}}=E^K_{n,-\vec{k}}\quad,\quad
\Phi^{K'}_{i,n,\vec{k}}(\vec{r})=\Phi^{K,*}_{i,n,-\vec{k}}(\vec{r}).
\eea
The Green's functions appearing in the Eq. \pref{SI:linearized_gap_eq}
are expressed by their spectral decomposition:
\bea
\mathcal{G}^{K/K'}_{ij}\left(\vec{r},\vec{r}';i\omega_l\right)&=&
\sum_{n\vec{k}}\frac{\Phi^{K/K'}_{i,n,\vec{k}}(\vec{r})\Phi^{{K/K'},*}_{j,n,\vec{k}}\left(\vec{r}'\right)}
{i\hbar\omega_l-\xi^{K/K'}_{n,\vec{k}}},
\eea
where: $\xi^{K/K'}_{n,\vec{k}}=E^{K/K'}_{n,\vec{k}}-\mu$ and $\mu$ is the chemical potential.
Using the Eq. \pref{SI:C2T_def}, we can write $\mathcal{G}^{K'}$
in terms of the eigenfunctions and energies at $K$:
\bea
\mathcal{G}^{K'}_{ij}\left(\vec{r},\vec{r}';i\omega_l\right)&=&
\sum_{n\vec{k}}\frac{\Phi^{K,*}_{i,n,\vec{k}}(\vec{r})\Phi^K_{j,n,\vec{k}}\left(\vec{r}'\right)}
{i\hbar\omega_l-\xi^K_{n,\vec{k}}}.
\eea
Because:
$\mathcal{G}^{K/K'}_{ij}\left(\vec{r},\vec{r}';i\omega_l\right)=
\mathcal{G}^{K/K'}_{ij}\left(\vec{r}+\vec{R}_m,\vec{r}'+\vec{R}_m;i\omega_l\right)$
for any moir\'e lattice vector $\vec{R}_m$, the solution of the linearized gap equation
\pref{SI:linearized_gap_eq} is periodic in the moir\'e unit cell:
\bea
\Delta^i(\vec{r})=\sum_{\vec{G}}\Delta^i\left(\vec{G}\right)e^{i\vec{G}\cdot\vec{r}}.
\eea
This implies that the Eq. \pref{SI:linearized_gap_eq} can be written as:
\bea\label{SI:linearized_gap_eq_TBG}
\Delta^i(\vec{r})&=&J\sum_{\vec{k}mn}
\frac{f\left(-\xi^K_{n,\vec{k}}\right)-f\left(\xi^K_{m,\vec{k}}\right)}
{\xi^K_{n,\vec{k}}+\xi^K_{m,\vec{k}}}
\Phi^K_{i,m,\vec{k}}(\vec{r})\Phi^{K,*}_{i,n,\vec{k}}(\vec{r})
\int\,d\vec{r}'\sum_j  
\Phi^{K,*}_{j,m,\vec{k}}\left(\vec{r}'\right)
\Phi^K_{j,n,\vec{k}}\left(\vec{r}'\right)
\Delta^j\left(\vec{r'}\right),
\eea
where we have performed the Matsubara sum
and introduced the Fermi distribution: $f(x)=\left(e^{\beta x}+1\right)^{-1}$.
We can safely assume that the terms with
$m\ne n$ barely contribute to the rhs of the Eq. \pref{SI:linearized_gap_eq_TBG},
and write:
\bea\label{SI:linearized_gap_eq_TBG_symp}
\Delta^i(\vec{r})&\simeq&J\sum_{\vec{k}m}
\frac{\tanh\left(\beta\xi^K_{m,\vec{k}}/2\right)}{
2\xi^K_{m,\vec{k}}}
\left|\Phi^K_{i,m,\vec{k}}(\vec{r})\right|^2
\int\,d\vec{r}'\sum_j  
\left|\Phi^K_{j,m,\vec{k}}\left(\vec{r}'\right)\right|^2
\Delta^j\left(\vec{r'}\right).
\eea
The Eq. \pref{SI:linearized_gap_eq_TBG_symp}
is equivalent to the following equation for the Fourier amplitudes,
$\Delta^i\left(\vec{G}\right)$'s:
\bea\label{SI:linearized_gap_eq_TBG_rec_space}
\Delta^i\left(\vec{G}\right)=\sum_{j\vec{G}'}\Gamma^{ij}\left(\vec{G},\vec{G}'\right)
\Delta^j\left(\vec{G}'\right),
\eea
where:
\bea
\Gamma^{ij}\left(\vec{G},\vec{G}'\right)\equiv \frac{J}{\Omega}\sum_{\vec{k}m}
\frac{\tanh\left(\beta\xi^K_{m,\vec{k}}/2\right)}{
2\xi^K_{m,\vec{k}}}
W^i_{m,\vec{k}}\left(\vec{G}\right)W^{j,*}_{m,\vec{k}}\left(\vec{G}'\right),
\eea
and:
\bea
W^i_{m,\vec{k}}\left(\vec{G}\right)\equiv\sum_{\vec{G}''}
\phi^{K,*}_{i,n,\vec{k}}\left(\vec{G}''\right)
\phi^K_{i,n,\vec{k}}\left(\vec{G}''+\vec{G}\right).
\eea
The critical temperature, $T_c$,
is obtained as the value of $T$ such that the largest eigenvalue of
the hermitian kernel $\Gamma$ is equal to $1$.
The Fourier transform of the corresponding eigenvector, $\Delta^i\left(\vec{G}\right)$,
provides the profile of the SC OP in the real space, $\Delta^i(\vec{r})$.
Because we consider values of the electronic density close to the CNP,
we take into account only the contribution of the two central bands
to the Eq. \pref{SI:linearized_gap_eq_TBG_rec_space}.

\begin{figure}
\includegraphics[width=3.in]{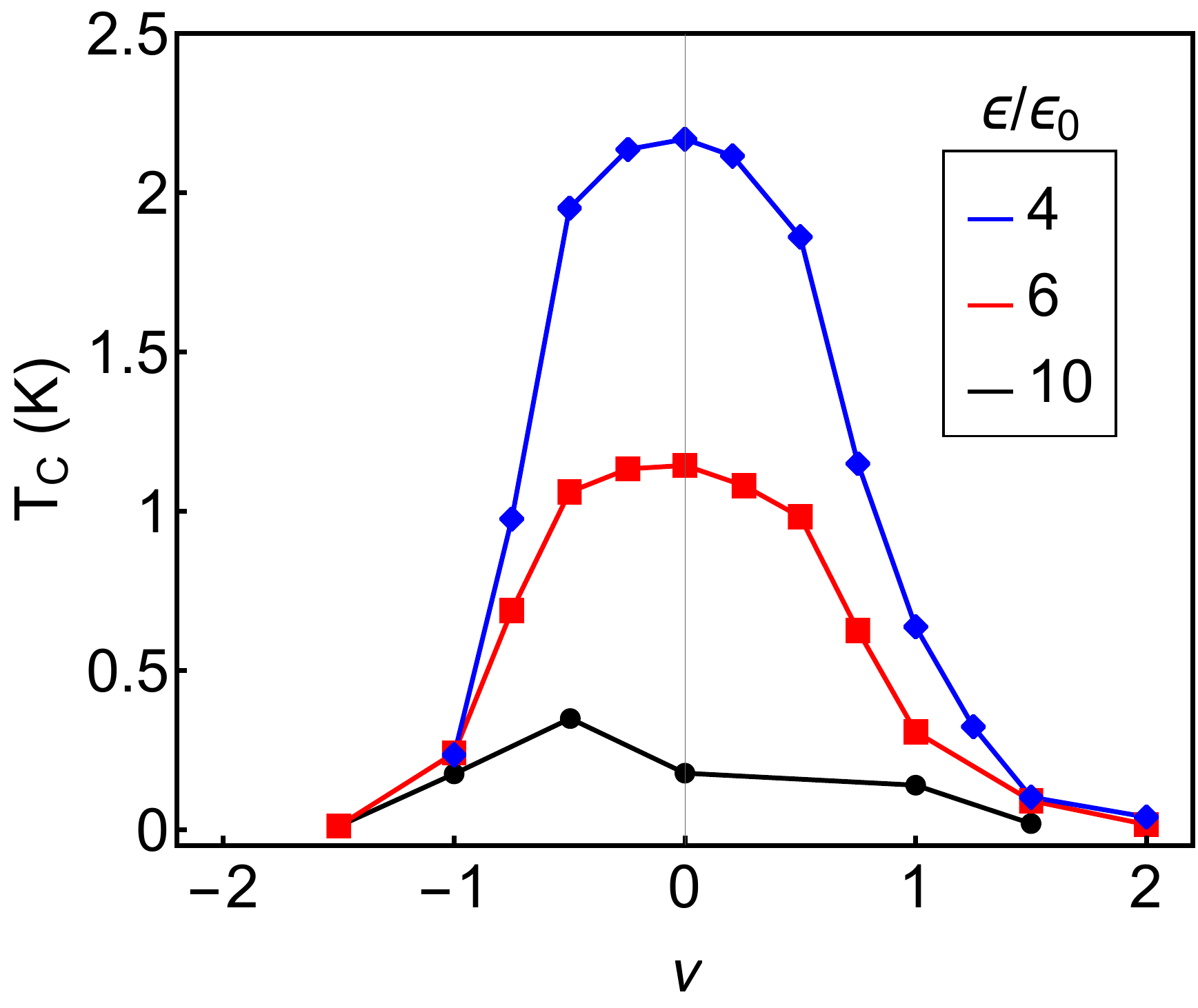}
\caption{
$T_c$ as a function of the filling per moir\'e unit cell, $\nu$,
obtained for: $\theta=1.05^\circ$ and
$\epsilon/\epsilon_0=4,6,10$.
}
\label{SI:fig:TBG_1p05deg_TC}
\end{figure}

The Fig. \ref{SI:fig:TBG_1p05deg_TC}
shows the values of $T_c$ as a function of the filling per moir\'e unit cell, $\nu$,
obtained for: $\theta=1.05^\circ$, which is close to a magic angle,
and $\epsilon/\epsilon_0=4,6,10$, as coded in the inset panel.
For the calculation,
we have used the parametrization of the TBG given by the
Ref. \cite{koshino_prx18}
and included the Hartree corrections to both the bands and the eigenfunctions
(see \cite{Guinea_pnas18,Cea_prb19}).
For the realistic values: $\epsilon/\epsilon_0\sim 4-6$,
$T_c$ is of the order of 1K
and its value increases with the DOS at the Fermi level, $N_F$,
explaining why we find the maximum of $T_c$ at $\nu=0$,
where the bandwidth is minimum (see the Fig. 2 of the manuscript).

\begin{figure}
\includegraphics[width=3.5in]{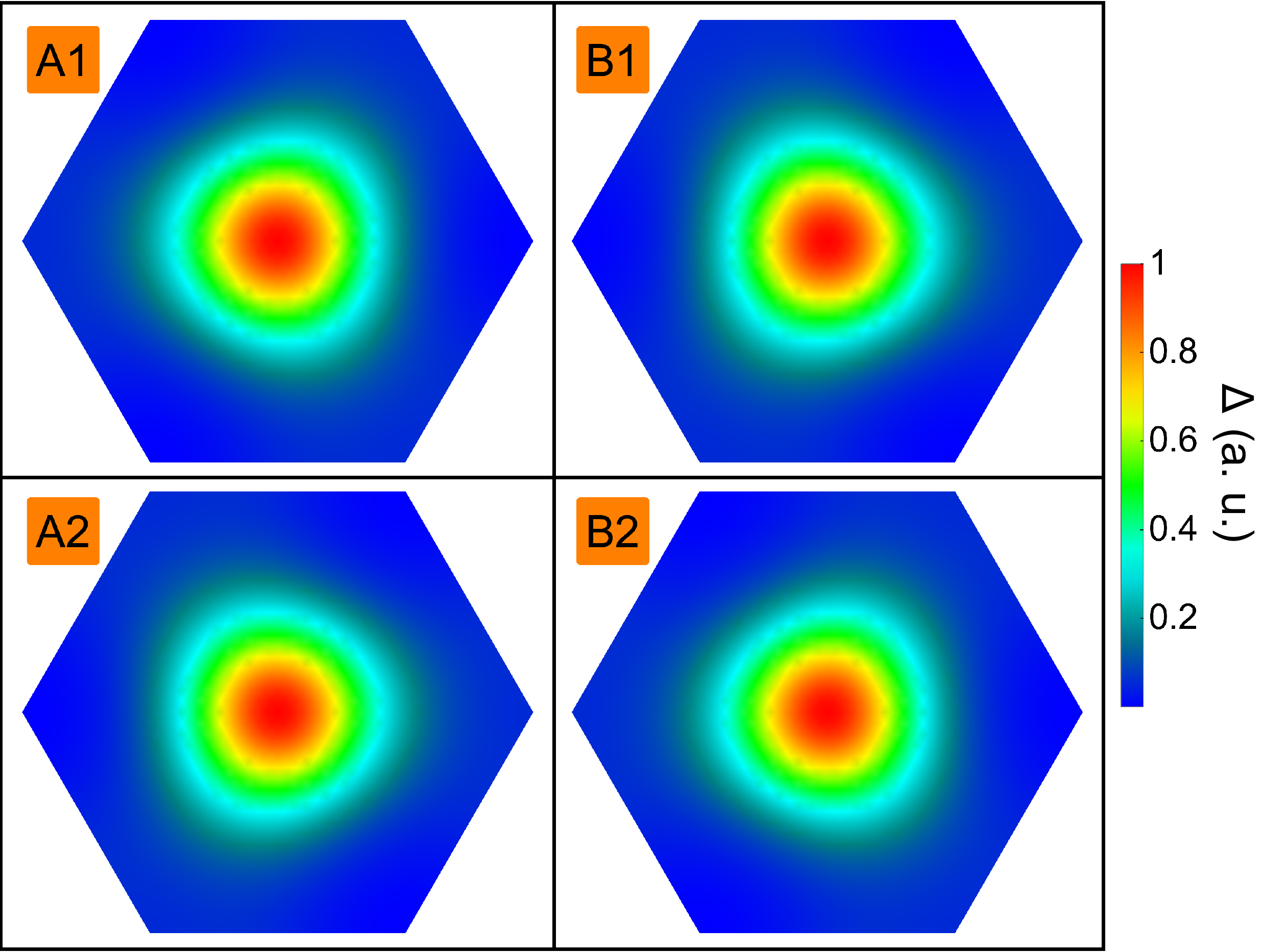}
\caption{
Profile of the SC OP
in the moir\'e unit cell, obtained for:
$\theta=1.05^\circ$, $\epsilon/\epsilon_0=4$ and $\nu=-0.5$.
Each panel refers to a different sub-lattice/layer combination,
where: $A,B$ identify the sub-lattice and $1,2$ the layer.
}
\label{SI:fig:OP_real_space_TBG}
\end{figure}

The Fig. \ref{SI:fig:OP_real_space_TBG} shows the profile of the SC OP
in the moir\'e unit cell, obtained for:
$\theta=1.05^\circ$, $\epsilon/\epsilon_0=4$ and $\nu=-0.5$.
Each panel refers to a different sub-lattice/layer combination,
where: $A,B$ identify the sub-lattice and $1,2$ the layer.
We find that the phase of the OP is always constant,
meaning that $\Delta^i(\vec{r})$ can be taken as real.
It's worth noting that the OP is well localized in the center of the unit cell,
the $AA$-stacked region, which is consistent with the distribution of the charge density
in the TBG\cite{bistritzer_pnas11,dossantos_prb12}.

\section{The case of the RTG and BBG}
\begin{figure}
\includegraphics[width=3.in]{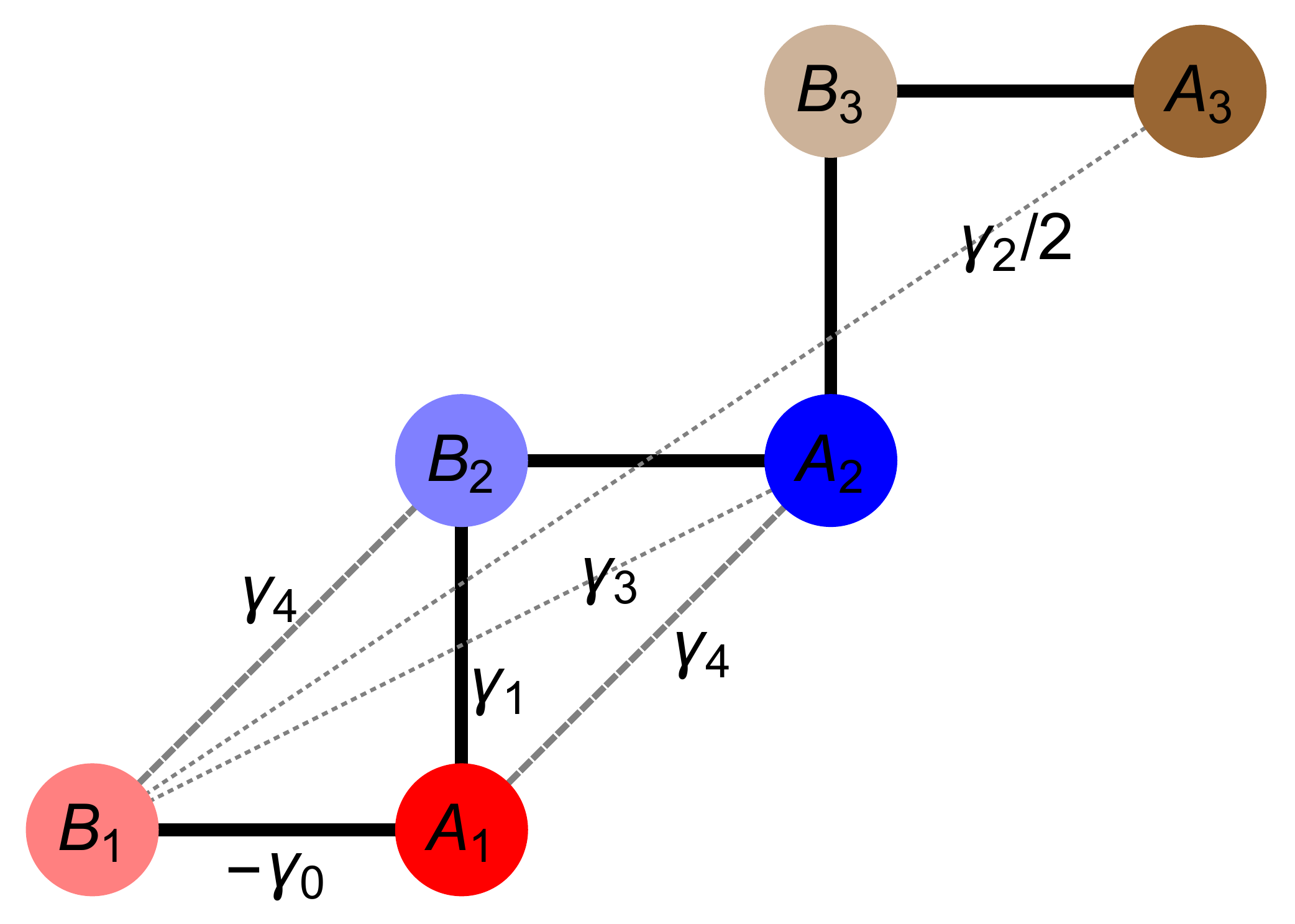}
\caption{
Schematic representation of the stacking arrangement
and of the hoppings of the $ABC$-stacked RTG.
The $\gamma_i$'s are the hopping amplitudes.
}
\label{SI:fig:ABC_scheme}
\end{figure}

We consider the ABC-stacked RTG and the AB-stacked BBG.
The tight binding model of the RTG can be described schematically
by the Fig. \ref{SI:fig:ABC_scheme},
where the $\gamma_i$'s are the hopping amplitudes.
A similar scheme can be sketched for the BBG,
where the outermost ($3^\mathrm{rd}$) layer is missing.
We use the low energy continuum model,
obtained by expanding the tight-binding Hamiltonian close to the points $K,K'$.
In the momentum space,
the Hamiltonian of the RTG is the $6\times6$ matrix in the basis
$(A_1,B_1,A_2,B_2,A_3,B_3)$:
\bea\label{SI:RTG_H_mat}
H_{K}(\vec{k})=
\begin{pmatrix}
\Delta_1+\Delta_2&v_0\pi(\vec{k})&-v_4\pi^*(\vec{k})&\gamma_1&0&0\\
v_0\pi^*(\vec{k})&\Delta_1+\Delta_2+\delta&-v_3\pi(\vec{k})&-v_4\pi^*(\vec{k})&
\gamma_2/2&0\\
-v_4\pi(\vec{k})&-v_3\pi^*(\vec{k})&-2\Delta_2&v_0\pi(\vec{k})&
-v_4\pi^*(\vec{k})&\gamma_1\\
\gamma_1&-v_4\pi(\vec{k})&v_0\pi^*(\vec{k})&-2\Delta_2&-v_3\pi(\vec{k})&
-v_4\pi^*(\vec{k})\\
0&\gamma_2/2&-v_4\pi(\vec{k})&-v_3\pi^*(\vec{k})&\Delta_2-\Delta_1+\delta&
v_0\pi(\vec{k})\\
0&0&\gamma_1&-v_4\pi(\vec{k})&v_0\pi^*(\vec{k})&\Delta_2-\Delta_1
\end{pmatrix},H_{K'}(\vec{k})=H^*_{K}(-\vec{k}),
\eea
where $v_i=\gamma_ia\sqrt{3}/(2\hbar)$, $\pi(\vec{k})=\hbar(k_x-ik_y)$,
$\Delta_2$ is the potential difference between the middle layer compared to
mean potential of the outer layers,
$\delta$ encodes an on-site potential which is only present
at sites $B_1$ and $A_3$ since these two atoms
do not have a neighbor on the middle layer and $\Delta_1$ is an interlayer bias,
which describes the effect of a displacement field perpendicular to the graphene's flakes
and opens a gap at the CNP.
The Hamiltonian of the BBG is given by the first $4\times4$ block of the Hamiltonian of the Eq. \pref{SI:RTG_H_mat}.  
We use the values of the $\gamma_i$'s, $\Delta_2$ and $\delta$
given by the Refs. \cite{zibrov_prl18,Zhou_nature21_bis,Cea_prb22},
that we report in the Table \ref{SI:table:ABC_parameters}.
\begin{table}
\begin{center}
\begin{tabular}{| c c c c c c c|}
\hline
$\gamma_0$&$\gamma_1$&$\gamma_2$&$\gamma_3$&$\gamma_4$&$\delta$&$\Delta_2$\\
\hline
3.1		    &0.38              &-0.015           &0.29              &0.141            &-0.0105 &-0.0023\\
 \hline
\end{tabular}
\end{center}
\caption{Minimal tight-binding parameters of the RTG and BBG.} 
\label{SI:table:ABC_parameters}
\end{table}

Exploiting the translational invariance, we consider the linearized gap equation
\pref{SI:linearized_gap_eq_TI}, with:
\bea\label{SI:RTG_greens_func}
\mathcal{G}^{K/K'}_{ij}(\vec{q},i\omega_l)=
\sum_{n}\frac{u^{K/K'}_{i,n,\vec{q}}u^{K/K',*}_{i,n,\vec{q}}}
{i\hbar\omega_l-\xi^{K/K'}_{n,\vec{q}}},
\eea
where $\vec{u}^{K/K'}_{n,\vec{q}}$
is the eigenvector of the Hamiltonian of the
Eq. \pref{SI:RTG_H_mat} at $\vec{q}$,
and: $\vec{u}^{K'}_{n,\vec{q}}=\vec{u}^{K,*}_{n,-\vec{q}}$.
Replacing the Eq. \pref{SI:RTG_greens_func} into the Eq. \pref{SI:linearized_gap_eq_TI}
and performing the Matsubara sum, we obtain:
\bea
\Delta^i=\sum_j\Gamma_{ij}\Delta^j,
\eea
where:
\bea
\Gamma_{ij}\equiv 
\frac{J}{\Omega}\sum_{\vec{q}nm}
u^{K}_{i,n,\vec{q}}u^{K,*}_{i,m,\vec{q}}
u^{K,*}_{j,n,\vec{q}}u^{K}_{j,m,\vec{q}}
\frac{f\left(-\xi^K_{m,\vec{q}}\right)-f\left(\xi^K_{n,\vec{q}}\right)}
{\xi^K_{n,\vec{q}}+\xi^K_{m,\vec{q}}}.
\eea
Because we consider values of the electronic density such that only one
of the two central bands crosses the Fermi surface
and the other bands are far away from the Fermi level,
we can safely assume that only the states in the partially filled band, $\bar{n}$,
contribute to the kernel, $\Gamma_{ij}$, and write:
\bea
\Gamma_{ij}\simeq
\frac{J}{\Omega}\sum_{\vec{q}}
\left|u^{K}_{i,\bar{n},\vec{q}}\right|^2
\left|u^{K}_{j,\bar{n},\vec{q}}\right|^2
\frac{\tanh\left(\beta\xi^K_{\bar{n},\vec{q}}\right)}{2\xi^K_{\bar{n},\vec{q}}}.
\eea

As we have described in the previous section for the case of the TBG,
we look for $T_c$ as the value of $T$ such that the largest eigenvalue of $\Gamma$ is equal to $1$.

Finally, we find that,
in the hole-doped regime analyzed in the Fig. 3 of the manuscript,
the SC OP is fully localized in the $A_3$ sub-lattice of the RTG
and in the $A_2$ sub-lattice of the BBG.
This is a direct consequence of what shown in the Ref. \cite{Zhang_prb10},
that the lowest energy states of the gapped RTG (BBG)
are fully localized on the $A_3$ ($A_2$) or on the $B_1$ sub-lattice,
depending if the system is hole- or electron-doped, respectively.


\section{Screening of the Coulomb interaction}
Here we estimate the screening of the Coulomb interaction by the particle-hole excitations
within the Random Phase Approximation (RPA).
We consider in particular the case of the RTG, which is translationally invariant,
but the arguments can be generalized to the other systems.
We consider both the long- and the short-range parts of the interaction.

The bare long-range interaction is described by the potential:
\bea
v_{LR}(\vec{q})=\frac{e^2}{2\epsilon |\vec{q}|}\text{ , for }\vec{q}\simeq 0.
\eea
Note that $v_{LR}$ does not depend on the internal indices of sub-lattice/layer, since the long-range interaction is an effective charge-charge repulsion.
The screening is given by the usual Thomas-Fermi formula:
\bea
\tilde{v}_{LR}(\vec{q})=\frac{1}{v^{-1}_{LR}(\vec{q})-\Pi(\vec{q})}\simeq -\frac{1}{\Pi(0)},
\eea 
where $\Pi(\vec{q})$ is the charge susceptivity that, at $\vec{q}=0$, can be written as:
\bea
\Pi(0)=\frac{4}{\Omega}\sum_{\vec{k}nm} \left|\vec{u}^K_{m,\vec{k}}\cdot \vec{u}^K_{n,\vec{k}}\right|^2
\frac{f\left(\xi^K_{n,\vec{k}}\right)-f\left(\xi^K_{m,\vec{k}}\right)}
{\xi^K_{n,\vec{k}}-\xi^K_{m,\vec{k}}},
\eea
where the factor of $4$ comes from the spin/valley degeneracy.
The leading contribution to $\Pi(0)$ is the intra-band contribution corresponding to the band crossing the Fermi level, $\bar{n}$,
so we can write:
\bea
\Pi(0)\simeq  \frac{4}{\Omega}\sum_{\vec{k}}
f'\left(\xi^K_{\bar{n},\vec{k}}\right)\begin{matrix}\\\to \\T\to0\end{matrix} -N_F.
\eea
If the Fermi level matches the VHS, then $N_F\to \infty$ and hence $\tilde{v}_{LR}(\vec{q})\to0$.

To derive the screening of the short-range interaction, namely of $J$,
it is convenient to consider the Hamiltonian of the Eq. \pref{SI:Hexc_cont}
and write the corresponding euclidean action:
\bea\label{S_exc}
S_{exc}=J
\sum_{i\sigma\sigma'}
\int\,dx
\psi^{K,\dagger}_{i\sigma}(x)
\psi^{K'}_{i\sigma}(x)
\psi^{K',\dagger}_{i\sigma'}(x)
\psi^{K}_{i\sigma'}(x),
\eea
where $x=(\vec{r},\tau)$ and $\tau$ is the imaginary time.
When computing the partition function,
$S_{exc}$ can be decoupled by introducing a set of complex Hubbard-Stratonovich (HS) fields,
$\Phi_i$, according to:
\bea
e^{-\frac{1}{\hbar}S_{exc}}&=&
\int\prod_i\mathcal{D}\left[\Phi_i,\Phi^*_i\right]
\exp\left\{
-\frac{\hbar}{J}\sum_i\int\,dx \left|\Phi_i(x)\right|^2+
i\sum_{i\sigma}\int\,dx\left[
\bar{\psi}^K_{i\sigma}(x)\psi^{K'}_{i\sigma}(x)\Phi_i(x)+c.c.
\right]
\right\}=\nn\\
&=&
\int\prod_i\mathcal{D}\left[\Phi_i,\Phi^*_i\right]
\exp\left\{
\frac{\hbar}{J}\sum_{iq}\left|\Phi_i(q)\right|^2+
i\sqrt{\frac{K_BT}{\hbar\Omega}}\sum_{i\sigma kk'}\left[
\bar{\psi}^K_{i\sigma}(k)\psi^{K'}_{i\sigma}(k')\Phi_i(k-k')+c.c.
\right]
\right\},\label{HS_transf}
\eea 
where
$\psi,\bar{\psi}$ are grassmanian numbers,
 $q=\left(\vec{q},i\Omega_l\right)$, $\Omega_l=2\pi l K_BT/\hbar$ are bosonic Matsubara frequencies,
and $k=\left(\vec{k},i\omega_l\right)$.
The Eq. \pref{HS_transf} defines the following self-energy:
\bea
\Sigma^{ij}_{\sigma\sigma'}\left(k-k'\right)\equiv
-i\delta_{ij}\delta_{\sigma\sigma'}\sqrt{\frac{K_BT}{\hbar\Omega}}
\begin{pmatrix}
0&\Phi_i\left(k-k'\right)\\
\Phi^*_i\left(k'-k\right)&0
\end{pmatrix},
\eea
where the $2\times2$ matrix acts in the valley space.
Integrating out the Fermionic fields, we obtain the effective action for the HS fields that, at the gaussian level,
is given by:
\bea
\frac{1}{\hbar}S_{eff}\left[\Phi,\Phi^*\right]&=&
\frac{\hbar}{J}\sum_{iq}\left|\Phi_i(q)\right|^2+
\frac{\hbar^2}{2}\sum_{kq}\mathrm{Tr}\left\{
\mathcal{G}(k+q)\Sigma(q)\mathcal{G}(k)\Sigma(-q)
\right\}=\nn\\
&=&
\hbar\sum_{ijq}\tilde{J}^{-1}_{ij}(q)\Phi^*_i(q)\Phi_j(q),
\eea 
where:
\bea
\tilde{J}^{-1}_{ij}(q)\equiv \frac{\delta_{ij}}{J}-2\frac{K_BT}{\Omega}\sum_k
\mathcal{G}^K_{ij}(k+q)\mathcal{G}^{K'}_{ji}(k)
\eea
is the inverse of the screened $J$. Note that $\tilde{J}$ is a matrix in the sub-lattice/layer indices.
This is not surprising, since $J$ is a matrix too, but proportional to the identity.
We focus on the $q=0$ limit. Using the spectral decomposition of the Green's functions $\mathcal{G}^K,\mathcal{G}^{K'}$
and performing the Matsubara sums,
we finally obtain:
\bea\label{J_scr}
\tilde{J}^{-1}_{ij}\equiv \tilde{J}^{-1}_{ij}(0)=
 \frac{\delta_{ij}}{J}-
 \frac{2}{\Omega}\sum_{\vec{k}nm}
 \left[u^K_{ni}(\vec{k}) u^K_{mi}(-\vec{k})\right]
 \left[u^{K}_{nj}(\vec{k}) u^K_{mj}(-\vec{k})\right]^*\times
 \frac{f\left(\xi^K_{m,-\vec{k}}\right)-f\left(\xi^K_{n,\vec{k}}\right)}
 {\xi^K_{m,-\vec{k}}-\xi^K_{n,\vec{k}}}.
\eea
Note that, since generally $\xi^K_{n,-\vec{k}}\ne\xi^K_{n,\vec{k}}$,
the intra-band contribution to the rhs of the Eq. \pref{J_scr} is not expected to provide a relevant weight
and, consequently, the strength of the short-range interaction is not expected to be strongly screened
as compared to its bare value, $J$.
This is a consequence of the fact that, at the level of the RPA, the processes contributing to screen $J$
are particle-hole excitations with the particle and the hole propagating in opposite valleys.

In order to give a quantitative estimate of the screening effects, we consider the RTG with $\Delta_1=50$meV
and $n_e=-0.75\times10^{12}$cm$^{-2}$. For this choice of the parameters, the Fermi level matches the VHS.
Using a very small temperature, $T=10^{-6}$K, we obtain:
\bea
\tilde{v}_{LR}(0)=0.01\text{eV{\AA}}^2\quad,\quad
\tilde{J}=\begin{pmatrix}
13.22&0&0&0.02&0&0\\
0&13.16&0&0&-0.01&0\\
0&0&13.07&0&0&0.01\\
0.02&0&0&13.22&0&0\\
0&-0.01&0&0&9.43&0\\
0&0&0.01&0&0&13.22
\end{pmatrix}\text{eV{\AA}}^2.
\eea
This means that: i) the long-range interaction is almost completely screened,
resulting order of magnitudes smaller than $J$.
Then one can assume that the long-range effects are not relevant.
ii) $\tilde{J}$ is approximately diagonal and its diagonal elements barely daviates
from the bare value: $J=13.25$eV{\AA}$^2$,
which allows one to safely neglect the screening of the short-range interaction, assuming:
$\tilde{J}_{ij}\simeq J\delta_{ij}$.

\newpage
\appendix
\def\appendixname{Appendix}

\section{}
\label{SI:app:V_DK_symm}
\begin{figure}[h!]
\includegraphics[width=4.in]{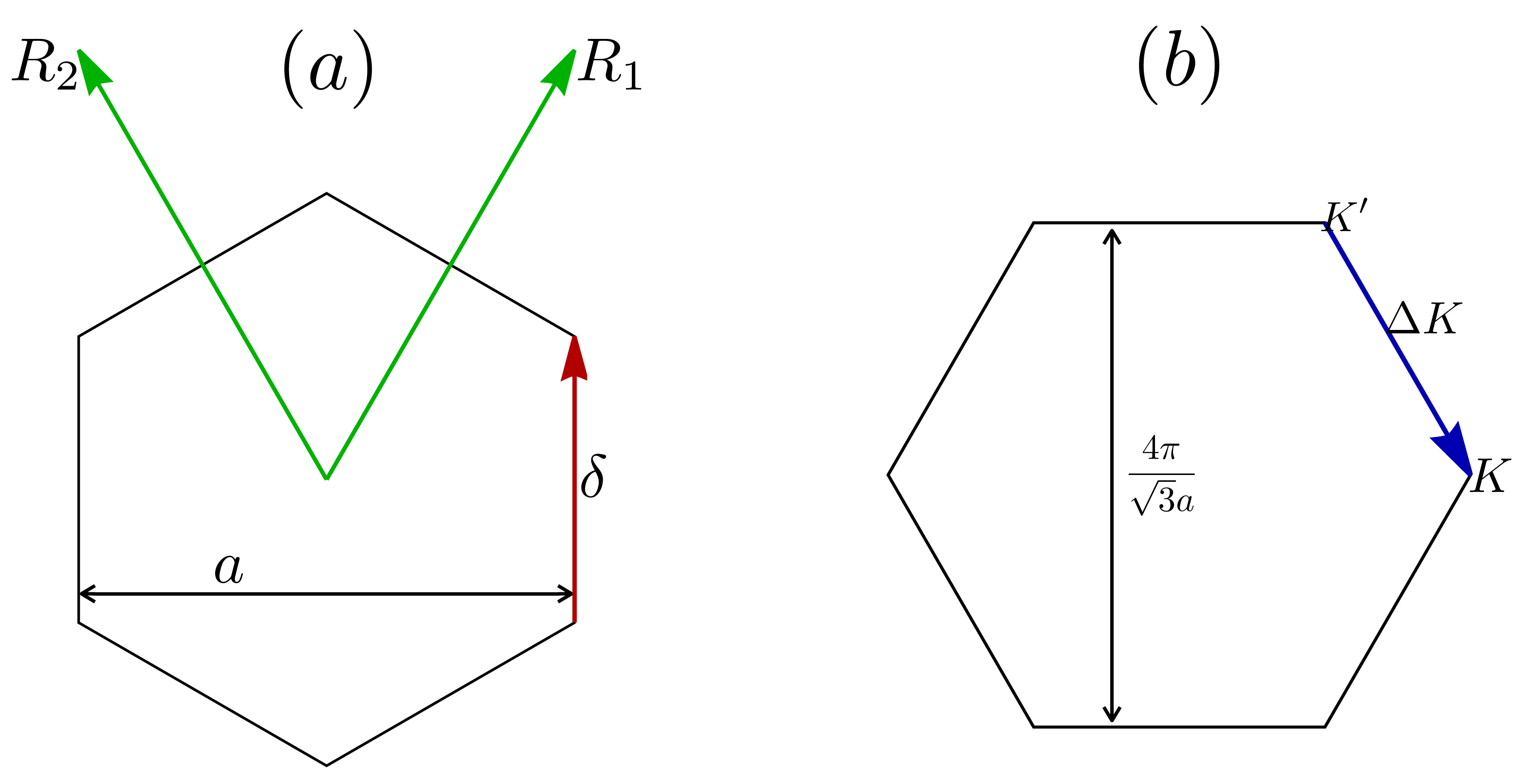}
\caption{(a) Unit cell of the monolayer graphene.
$\vec{R}_1$ and $\vec{R}_2$ are the unit vectors of the Bravais lattice
and $\vec{\delta}$ is the atomic displacement between the two sub-lattices.
(b) Brillouin zone of the monolayer graphene.
$K,K'$ are the two non equivalent corners and $\Delta K=K-K'$.
}
\label{SI:fig:uc_BZ}
\end{figure}

We show that, if $i,j$ correspond to the same layer but to different sub-lattices, then:
\bea\label{SI:statement_V=0}
\sum_{\vec{R}}V^{ij}_{C}(\vec{R})e^{-i\Delta K\cdot\vec{R}}\propto
\sum_{\vec{R}}\frac{e^{-i\Delta K\cdot\vec{R}}}{\left|\vec{R}+\vec{\delta}_i-\vec{\delta}_j\right|}=0,
\eea
where: $\vec{R}=n_1\vec{R}_1+n_2\vec{R}_2$, with $n_1,n_2$ integer,
and $\vec{R}_1=a(1,\sqrt{3})/2$, $\vec{R}_2=a(-1,\sqrt{3})/2$
are the two unit vectors of the Bravais lattice of the monolyer graphene, as shown in the Fig. \ref{SI:fig:uc_BZ}(a).
Without loss of generality, we can assume that the atomic displacement between the two sub-lattices
in each unit cell is:
$\vec{\delta}\equiv\vec{\delta}_i-\vec{\delta}_j=a(0,1)/\sqrt{3}$, and that:
$\Delta K=2\pi(1,-\sqrt{3})/(3a)$ (see the Fig. \ref{SI:fig:uc_BZ}(b)).

Considering that, for any $n$ integer,
the Bravais lattice is invariant under the rotation by $2n\pi/3$, $C_n$,
we can write the sum in the Eq. \pref{SI:statement_V=0} as:
\bea\label{SI:V_sum_step1}
\sum_{\vec{R}}\frac{e^{-i\Delta K\cdot\vec{R}}}{\left|\vec{R}+\vec{\delta}\right|}=
\frac{1}{3}\sum_{\vec{R}}\left[
\frac{e^{-i\Delta K\cdot\vec{R}}}{\left|\vec{R}+\vec{\delta}\right|}+
\frac{e^{-i\Delta K\cdot\left(C_1\vec{R}-\vec{R}_1\right)}}{\left|C_1\vec{R}-\vec{R}_1+\vec{\delta}\right|}+
\frac{e^{-i\Delta K\cdot\left(C_2\vec{R}-\vec{R}_2\right)}}{\left|C_2\vec{R}-\vec{R}_2+\vec{\delta}\right|}
\right],
\eea
where, in addition to the rotation $C_1(C_2)$,
we have performed a translation by $-\vec{R}_1(-\vec{R}_2)$.
Noting that:
\bea
\vec{\delta}-\vec{R}_{1,2}=C_{1,2}\vec{\delta}\quad\text{, and:}\quad
e^{-i\Delta K\cdot C_{1,2}\vec{R}}=e^{-i\Delta K\cdot\vec{R}},
\eea
then we can write the Eq. \pref{SI:V_sum_step1} as:
\bea
\sum_{\vec{R}}\frac{e^{-i\Delta K\cdot\vec{R}}}{\left|\vec{R}+\vec{\delta}\right|}=
\sum_{\vec{R}}\frac{e^{-i\Delta K\cdot\vec{R}}}{\left|\vec{R}+\vec{\delta}\right|}\times
\left(\frac{1+e^{i\Delta K\cdot\vec{R}_1}+e^{i\Delta K\cdot\vec{R}_2}}{3}\right)=0,
\eea
as it follows by:
$1+e^{i\Delta K\cdot\vec{R}_1}+e^{i\Delta K\cdot\vec{R}_2}=1+e^{-2i\pi/3}+e^{2i\pi/3}=0$.

\end{document}